%% file: main.tex
\documentclass[fleqn,10pt]{wlscirep}
\usepackage[utf8]{inputenc}
\usepackage[T1]{fontenc}
\usepackage{lineno}

\usepackage{algorithm}
\usepackage{algorithmic}
\usepackage{adjustbox}

\linenumbers

\usepackage{subfig}

\usepackage{dirtree}

\nolinenumbers

\title{Towards Unifying Anatomy Segmentation: Automated Generation of a Full-body CT Dataset via Knowledge Aggregation and Anatomical Guidelines}

\author[1,7$\dag$,*]{Alexander Jaus}
\author[1,3,$\dag$]{Constantin Seibold}
\author[3]{Kelsey Hermann}
\author[2,4,5,7]{Alexandra Walter}
\author[4,5]{Kristina Giske}
\author[6]{Johannes Haubold}
\author[3,7,**]{Jens Kleesiek}
\author[1,7,**]{Rainer Stiefelhagen}
\affil[1]{Institute for Anthropomatics and Robotics, Karlsruhe Institute of
Technology, Karlsruhe, Germany}
\affil[2]{Steinbuch Center for Computing, Karlsruhe Institute of Technology (KIT), Karlsruhe,
Germany}
\affil[3]{Institute for AI in Medicine, University Hospital Essen, Essen, Germany}
\affil[4]{Department of Medical Physics in Radiation Oncology, German Cancer Research Center (DKFZ), 
Heidelberg, Germany}
\affil[5]{Heidelberg Institute of Radiation Oncology (HIRO) \& National Center for Radiation Research in Oncology (NCRO), Heidelberg/Dresden, Germany}
\affil[6]{Department of Diagnostic and Interventional Radiology and Neuroradiology, University Hospital Essen, Essen, Germany}
\affil[7]{Helmholtz Information and Data Science School for Health (HIDSS4Health), Karlsruhe/Heidelberg, Germany}

\affil[*]{corresponding author(s): Alexander Jaus (alexander.jaus@kit.edu)}

\affil[$\dag$]{these authors contributed equally to this work}

\affil[**]{Shared last author}

\begin{abstract}
\input{Latex/Abstract}
\end{abstract}
\begin{document}

\flushbottom
\maketitle

\thispagestyle{empty}

\input{Latex/Background_Summary}
\input{Latex/Methods}

\input{Latex/Data_Records}
\input{Latex/Technical_Validation}

\input{Latex/Discussion}

\section*{Usage Notes}
The AutoPET dataset~\cite{gatidis2022whole} contains three different image types for each examination: The PET and the corresponding SUV, the tumor segmentation mask, and two different CT images. One of them is in the original resolution whereas the second one is resampled to match the PET/SUV resolution. DAP Atlas is based on the original CT image resolution, however, resampling of the masks can be performed to match the PET resolution for future multimodel PET/CT work. 

\section*{Code availability}

The code for the dataset aggregation and for the post-processing will be made publicly available under \href{https://github.com/alexanderjaus/AtlasDataset}{github}. 
Besides the dataset, we publish the trained models V1 as well as the robust model V2 at \href{https://github.com/alexanderjaus/AtlasDataset}{github}. 

\section*{Conflicts of Interest}
The authors disclose no conflicts of interest.

\section*{Informed Consent Statement:}
The DAP Atlas uses a subset of the AutoPET dataset. As the AutoPET dataset was part of a public competition, ethical approval from our side was not required, as confirmed by the license attached with the open-access data. 
\bibliography{main}


\section*{Acknowledgements} 
The present contribution is supported by the Helmholtz Association under the joint research school “HIDSS4Health – Helmholtz Information and Data Science School for Health”.
It was performed on the HoreKa supercomputer funded by the Ministry of Science, Research and the Arts Baden-Württemberg and by the Federal Ministry of Education and Research and is supported by the Helmholtz Association Initiative and Networking Fund on the HAICORE@KIT partition. 

\section*{Author contributions statement}
A.J., C.S.: conception, design and creation of the Atlas dataset and evaluation procedure. 
K.H., J.H.: Human Feedback of the dataset.
A.W., K.G.: data collection of private Head-Neck dataset and training of the expert model.  
J.K., R.S.: discussion and supervision of the project.

\end{document}

%% file: Latex/Abstract.tex
In this study, we present a method for generating automated anatomy segmentation datasets using a sequential process that involves nnU-Net-based pseudo-labeling and anatomy-guided pseudo-label refinement. By combining various fragmented knowledge bases, we generate a dataset of whole-body CT scans with $142$ voxel-level labels for $533$ volumes providing comprehensive anatomical coverage which experts have approved. Our proposed procedure does not rely on manual annotation during the label aggregation stage. We examine its plausibility and usefulness using three complementary checks: Human expert evaluation which approved the dataset, a Deep Learning usefulness benchmark on the BTCV dataset in which we achieve $85\%$ dice score without using its training dataset, and medical validity checks. This evaluation procedure combines scalable automated checks with labor-intensive high-quality expert checks.

Besides the dataset, we release our trained unified anatomical segmentation model capable of predicting $142$ anatomical structures on CT data.

%% file: Latex/Background_Summary.tex
\section*{Background \& Summary}
\label{Sec: Background}
Medical image segmentation has shown tremendous success ever since the advent of Deep Learning within the medical field which can be dated back to the introduction of the U-Net~\cite{ronneberger2015u}. While there has been a lot of architectural advancement of models to generalize U-Net architectures to the prevalent 3D medical imaging modalities~\cite{cciccek20163d,milletari2016v} such as Computed Tomography (CT), Magnetic Resonance Imaging (MRI) or  Positron Emission Tomography (PET), there has been less effort in providing a holistic view on the entire human body and its anatomy. 
Current models for medical data predominantly specialize in partially annotated datasets on sub-areas of the human body. 
These datasets range from single organ annotations such as the segmentation of spleen or pancreas~\cite{simpson2019large} to multi-organ datasets such as the BTCV~\cite{landman2015miccai} containing annotations for 13 abdominal organs. The recently introduced Medical Segmentation Decathlon~\cite{antonelli2022medical} aims at generalizing models to multiple tasks. Each of the individual tasks is however still limited to a specific body region and a certain organ of interest which does not address a complete anatomical view.  
Focusing on only a subsection of the entire anatomy has multiple downsides. From a technical point of view, it limits potential downstream applications which could be used within clinical settings. Additionally, it is a well-known fact that the human body consists of multiple systems and structures which are in constant interaction with each other to maintain a homeostatic state. Thus also from a medical point of view, it appears beneficial to look at interactions between structures instead of treating them in an isolated fashion. This requires overcoming the current constrained setting and striving towards a holistic view of anatomical structures which is a limitation of the current literature.

A common problem for Deep Learning applications is the need for large annotated datasets. The medical domain is no exception and poses additional difficulties due to the required expertise to annotate medical images combined with the labor-intensive annotation process for common 3D imaging modalities. While this is already cumbersome for a small number of labels, it is effectively impossible to find a team of radiologists willing to annotate the combination of hundreds of categories for hundreds of CTs. For example, the recently proposed ATM-dataset~\cite{zhang2023multi}, comparable in size with the proposed Dense Anatomical Prediction (DAP) Atlas dataset, utilizes three expert radiologists working on a single label with each CT volume requiring about 60-90 minutes~\cite{zhang2023multi}. Assuming an average time of $75$ minutes per CT volume, this single-label dataset requires almost $80$ days of work, assuming $8$h of uninterrupted work per day. Scaling to more labels is infeasible as it will not only increase the workload but also might cause the radiologist to lose track due to the higher complexity of the task. 

The creators of MOOSE~\cite{sundar2022fully}, a multi-organ segmentation model use a hybrid approach consisting of expert annotation and automated annotations. The experts annotate on a small dataset of $50$ CT images $13$ organ structures, $20$ bone segments, and $4$ tissue structures which were semi-automatically extracted. For the majority of their labels which are $83$ cerebral structures, the authors use an automatic segmentation approach by leveraging the Hammersmith atlas~\cite{hammers2003three}.

TotalSegmentator~\cite{wasserthal2022totalsegmentator}  approaches this problem via active learning in which an expert improves model predictions which are again fed to the model to improve the predictions. While this procedure noticeably reduces an expert's time spent on annotation, it still relies on direct interaction with experts for multiple weeks to generate a dataset of 104 anatomical structures. It also has to be acknowledged that large datasets with this many labels face a concrete challenge when evaluating the label quality since it becomes increasingly infeasible to have pixel-wise alignments checked for all anatomical structures. As such, TotalSegmentor~\cite{wasserthal2022totalsegmentator} relied upon quality insurance via 3D renderings instead of manual voxel-wise alignment checks. 

When we compare this to the natural image domain, several works utilize entirely automated annotation for classification and segmentation purposes\cite{xie2020self,zhai2022scaling,kirillov2023segment}. One example is the recent \textit{Segment Anything}-dataset~\cite{kirillov2023segment}. Here, the authors train a base model on manually annotated data and make predictions on unlabeled images through multi-scale inference and filter predictions via non-maximum suppression leading to 11 million annotated images. The general concept of pseudo-label filtering from weak annotations like image-level class labels improves the unlabeled training data and leads to more stable models\cite{reiss2023decoupled,wang2022omni,bae2022one,mlynarski2019deep,qu2023annotating}.  
Other works~\cite{hu2023label} have shown the successful application of pseudo-generated tumor labels on CT data of the liver leading to accurate segmentation results on real liver tumors.

Motivated by these recent advancements and the difficulty to scale expert annotations to multi-label large-scale medical datasets, we want to develop a dataset that enables the training of models suitable to serve a variety of clinically relevant downstream tasks which profit from extensive anatomical knowledge like body composition analysis, surgery planning or cancer treatment monitoring. 

\begin{figure}
    \centering
    \includegraphics[trim= 0 0 0 0, clip, width=\textwidth]{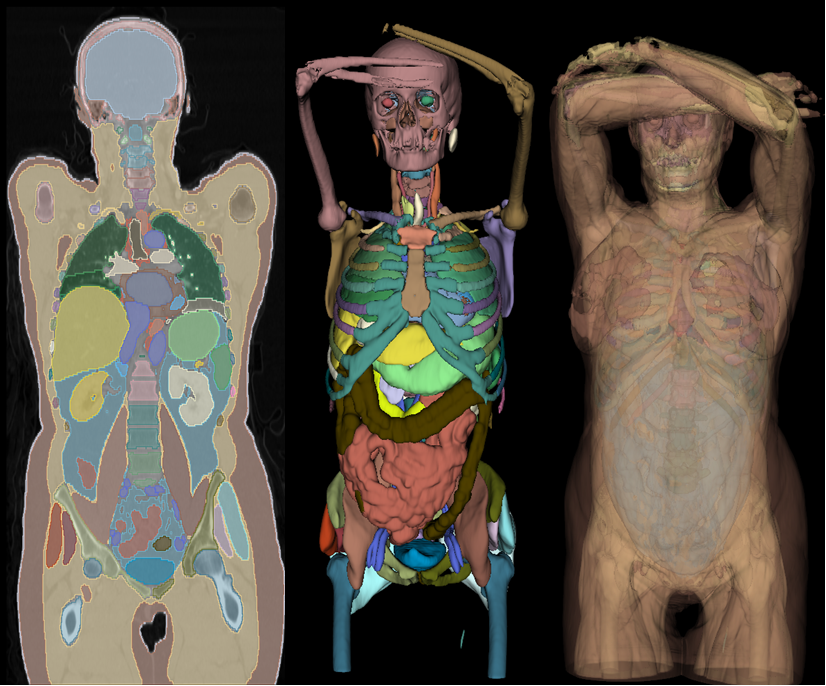}
    \caption{A sample image of the proposed Dense Anatomical Prediction Atlas dataset in a coronal slice next to two 3D images without tissues and including tissues rendered by 3DSlicer~\cite{slicer}.}
    \label{fig:intro_fig}
\end{figure}

We build upon this core idea of pseudo-label filtering and refinement to employ an expert-free dataset generation approach that aggregates the scattered anatomical knowledge of multiple source datasets and aggregates them into a single dataset. 
We combine anatomical information from various sources through pseudo-label-based label aggregation and pseudo-label-refinement in post-processing strategies that leverage anatomical textbook knowledge to assert the anatomical plausibility of the labels. 
Our dataset has been approved by experts, despite not having contributed to its creation.  
The dataset consists of $533$ whole-body CT images with labels for $142$ anatomical structures containing information ranging from body composition, organs to various vessels. We are the first to release a dataset containing dense annotations for every voxel in a full-body human CT. We show an example of the dataset in Fig.~\ref{fig:intro_fig}. 

The remainder of this paper is structured as follows: Section \textit{Methods} discusses how the full-body CT data and the labels are acquired and how they are processed to generate our Atlas dataset. In Section \textit{Data Records} we briefly discuss the structure and availability of the dataset to ease the usage for the community. This is followed by Section \textit{Technical Validation} in which the quality of the dataset is examined. We conclude the paper with the final chapter \textit{Discussion and Conclusion}. 

%% file: Latex/Methods.tex
\section*{Methods}
\label{S: Methods}
\subsection*{Data Acquisition}

Two of the most common volumetric modalities are CT and MRI. 
While MRI often focuses on soft tissue analysis and brain imaging, CT is a common choice in the clinical routine due to its acquisition time and broad field of use. As we aim to generate models to segment any anatomy utilizing various sources, we start by selecting a dataset that acts as a solid basis for full-body label aggregation.
The recently published AutoPET dataset~\cite{gatidis2022whole} is a PET-CT dataset that perfectly fits our requirements since nuclear medicine often requires full-body CT scans to track therapy. In addition to the full-body CTs, this dataset might enable future multi-modal segmentation tasks~\cite{xue2021multi, marinov2023mirror} due to the separate PET domain and lesion annotations. Future multimodal tasks could make use of the provided anatomical structures, which, however, is not the focus of this work.

We select a subset of $566$ CTs of the AutoPET dataset. The selection criterion is based on similar slice thickness in the axial dimension leading to a homogeneous dataset. Furthermore, we make sure that the images show important regions of interest. Our region of interest starts from the head and ends slightly below the hip which includes all thoracic and abdominal organs. The chosen subset consists of the CTs which contain between $336$ and $400$ slices in the original dicom files. We exclude CTs with fewer slices as these tend to show an insufficient subpart of the body contradicting the desired full-body dataset. CT images containing more than 400 slices tend to include more irrelevant content. 
This leaves us with a homogeneous dataset of size $566$. Our final DAP Atlas dataset does deviate from this selection, as we filter out implausible predictions, in a final post-processing step leaving us with $533$ CT images. The filter criteria will be described in detail later. In the following, we will refer to the DAP Atlas dataset as the dataset containing $533$ images. 

\begin{figure}
    \centering
    \includegraphics[trim= 105 200 120 100, clip, width=0.9\textwidth]{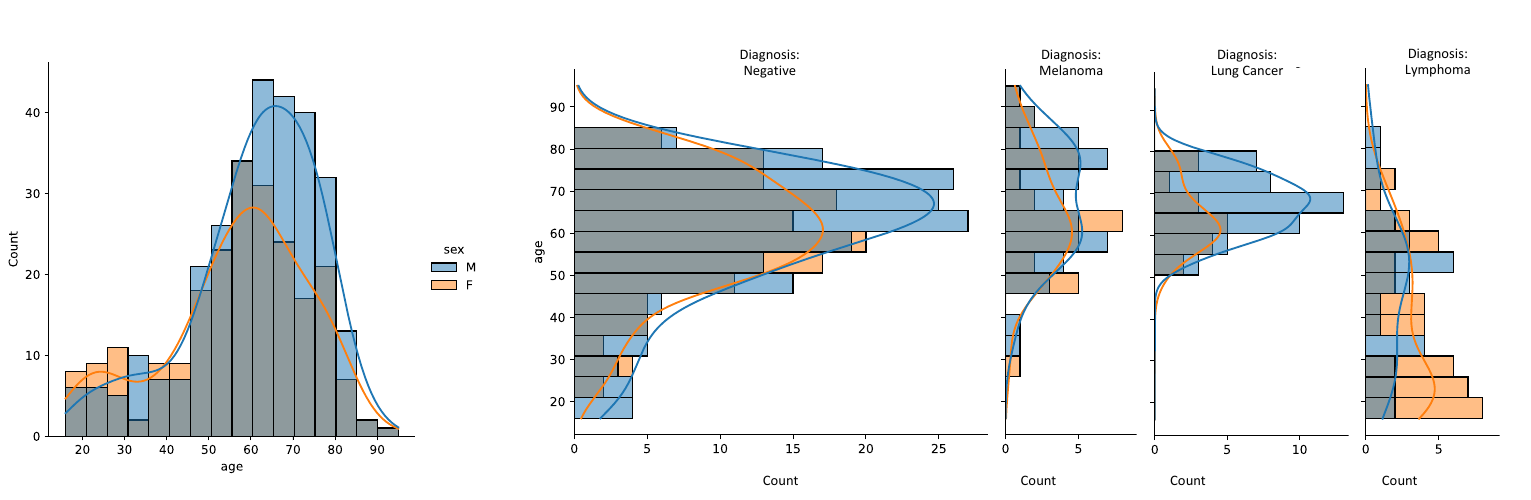}
    \caption{Descriptive statistics of the DAP Atlas dataset. On the left, we show the age distribution for men and women. The distribution peaks between $60$ and $70$ years which is common for medical datasets, as radiological examinations are usually performed after an initial suspicion of disease.
    The following four diagrams break down the distribution of the pathological findings by sex and by age.}
    \label{fig:descriptive_statistics}
\end{figure}

Our DAP Atlas is similar to AutoPET regarding the age and gender distributions as well as pathological findings. We show a descriptive analysis of the dataset regarding the aforementioned dimensions in Fig.~\ref{fig:descriptive_statistics}. 

\subsection*{Knowledge Acquisition}

 \begin{figure}[t]
     \centering
     \includegraphics[width=\textwidth]
     {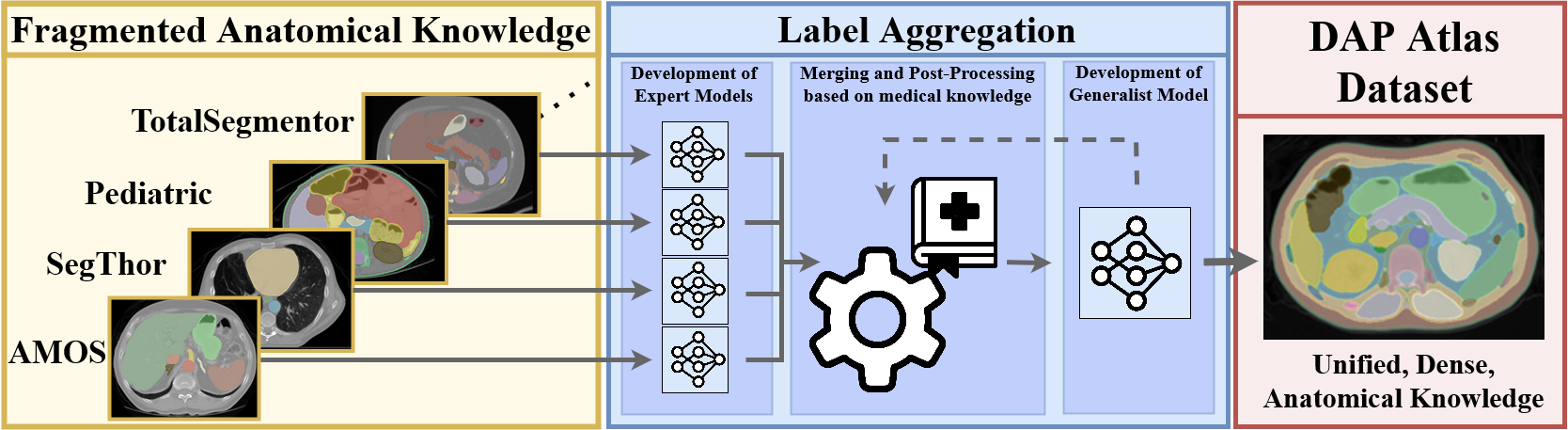}
     \caption{Overview of the creation of the DAP Atlas dataset. It combines several sources of fragmented anatomical knowledge. After training expert models on publicly available datasets, utilizing non-publicly available models, and deriving new labels from anatomical rules, we aggregated the different sources of knowledge and refine them using anatomical textbook knowledge. The resulting model is applied to the AutoPET dataset and the predictions are refined by a post-processing algorithm. After aggregation, we perform one iteration of self-training. The final DAP Atlas dataset is constructed by predicting the entire anatomy using a single model, these predictions are again refined by the post-processing algorithm. 
     This resulting dataset becomes the final DAP Atlas dataset. We thus gradually combine multiple sources of knowledge to provide a holistic, dense label map spanning $142$ anatomical labels.}
     \label{fig:merging}
 \end{figure}

The dataset aggregates multiple sources of anatomical segmentation knowledge which we differentiate into public knowledge which is present in the form of publicly available datasets and private knowledge which are private datasets available to us. Besides the segmentation knowledge, we leverage rule-based knowledge which is derived from anatomical textbook knowledge and represents what could be described as the common sense of a radiologist. These rules contain for instance, which anatomical structures are possible in which part of the human body. We display the DAP Atlas construction workflow in Fig.~\ref{fig:merging} and discuss the details in the following.

A large amount of labels of the dataset is derived from public segmentation knowledge which is present in fragmented form through publicly available datasets. These contain annotations of organs of interest on CT images showing parts of the human body. We extract this knowledge by training neural networks on these public datasets, which learn to predict the labels of the respective datasets. Typically, training a neural network is a data and task-specific problem and requires finetuning a large set of hyperparameters which is impracticable for our desired applications as we intend to train a vast number of models on various heterogeneous datasets. To overcome this problem, we use the nnU-Net~\cite{isensee2021nnu}, a framework that automatically configures a U-Net and adapts the training procedure to the data at hand. This Auto-ML framework provides segmentation results surpassing several more complex works without any of their additional engineering overhead. We employ standard nnU-Nets and train them on publicly available datasets. The used publicly available datasets are shown along with their obtained label category in Fig.~\ref{Tab:Label_Usage}. After training, these networks are used to carry over the extracted knowledge by predicting the learned labels into our full-body selected DAP Atlas CT images. We describe the used datasets and the merging procedure in the following. 

\input{tabel_labels}

\begin{itemize}
    \item \textbf{Pediatric~\cite{jordan2022pediatric}:} This dataset consists of $359$ chest-abdomen-pelvis and abdomen-pelvis CT images of patients between the age of $5$ and $16$ years. It provides $29$ anatomical structures annotated by experts. Patients were selected based on random clinical indications from the university clinic of Children's Wisconsin.
    \item \textbf{Total Segmentator~\cite{wasserthal2022totalsegmentator}:} The TotalSegmentator dataset is large and diverse with $1024$ CT images of different body parts with labels for $104$ anatomical structures. The dataset was collected by randomly sampling from the PACs systems of multiple sites. Its annotation is based on an interactive semi-automatic approach. Here, models are first trained on a few manual annotations. These models infer predictions on unlabeled scans which are lastly refined by an expert. This cycle repeats with an ever-increasing number of training images.
    \item \textbf{SegThor~\cite{lambert2020segthor}:} A dataset consisting of $60$ thoracic CTs collected at the Henri Becquerel Center. The patients were selected based on lung cancer or Hodkin's lymphoma diagnosis. The CTs contain annotations for four organs at risk whose tissues must remain intact during radiation therapy. The annotations of the dataset are provided by an experienced radiotherapist. 
    \item \textbf{CT50Abdomen~\cite{ma2021abdomenct}:} The dataset is part of the CT1k Abdomen datasets extension in which the authors provide 50 abdominal CT images with previously less annotated structures such as the adrenal glands. Annotations are provided by multiple junior annotators and checked by senior radiologists.
    \item \textbf{MAL Cervix~\cite{landman2015miccai}:} This dataset is part of the Beyond the Cranial Vault challenge. It consists of $30$ training and $20$ testing abdominal CT images acquired via a full bladder drinking protocol and annotated by a trained radiation oncologist. It focuses on the digestive and reproductive systems of female cervical cancer patients.
    \item \textbf{Amos~\cite{ji2022amos}:} A diverse dataset with $500$ CT images collected from different scanners and sites covering 15 abdominal organ categories. The selection of patients relates to abdominal tumors or abnormalities examinations. Annotations rely on a combination of junior and senior radiologist labor.
    \item \textbf{RibSeg~\cite{yang2021ribseg}:} The RibSeg dataset consists of $490$ CT Scans taken from publicly available RibFrac~\cite{jin2020deep} dataset. The authors use a semi-automatic morphology-based segmentation approach based on thresholding, point cloud segmentation, and morphological operations. They check the proposed segmentations by hand and refine them if necessary. 
    \item \textbf{Verse ~\cite{sekuboyina2021verse}:} A large dataset for vertebra segmentation. It consists of two subsets and has a total of $374$ CT scans of $355$ patients from multiple detectors and sites with voxel-wise annotations for individual vertebras. Segmentations have been performed semi-automatically with initial proposals being generated by an in-house pipeline. The proposals are refined by a team of trained medical students and experts and finally approved by a radiologist with more than $30$ years of experience. 
    \item \textbf{ATM~\cite{zhang2023multi}:} This dataset establishes a benchmark for Airway Tree Modelling by providing $500$ chest CT scans from different sites and includes scans of healthy patients, patients with pulmonary diseases, and even noisy COVID-19 CTs. Annotations of the pulmonary airways were performed by a team of three experts with each radiologist having more than five years of experience.
    \item \textbf{PARSE~\cite{kuanquan_wang_2022_6361906}:} The PARSE dataset is part of the Pulmonary Artery Segmentation Challenge and contains a total of $203$ CT images from $203$ patients which have been diagnosed with pulmonary nodular diseases. The CTs were generated using devices from two different manufacturers, with data collected from four distinct sites. Each of the images has been annotated by five experts with each expert having at least five years of experience in the field. 
    \item \textbf{Pelvic CT~\cite{liu2021deep}:} A large-scale dataset that focuses on the segmentation of pelvic bone structures such as hip bones or sacrum. It consists of $1184$ CT images collected from different source datasets combining images from multiple sites, scanners, and even metal artifacts. The labeling was conducted by a team of junior and senior radiologists.
    
\end{itemize}


Besides the previously described dataset, we leverage non-publicly available datasets and models. One of the models is the body composition analysis model~\cite{koitka2021fully} which differentiates between different types of tissues. From this model, we obtain labels such as \textit{fat} or the general class \textit{muscles}. In total, we extract $9$ labels from the body composition model source. 
A second private source dataset consisting of $104$ diverse head and neck contrast CT images from four different source cohorts~\cite{giske2011local, stoiber2017analyzing, bejarano2019longitudinal, bejarano2018head, clark2013cancer}.
This dataset focuses on the diagnosis and treatment of oropharyngeal or hypopharyngeal head and neck cancer and has been annotated by medical students. It provides fine-grained classes for head neck vessels and bone structures. We train a standard nnU-Net on $86$ images to extract the dataset knowledge and add $12$ unique, previously unavailable labels from this dataset, mostly vessels in the head-neck region. 

After obtaining the labels from the different nnU-Net predictions, we use anatomically derived rules to refine the current predictions and generate $7$ additional labels. An intuitive example for a new label that can be derived from the combination of obtained labels and medical common sense is the skull. It can be derived from a thresholding procedure obtained by the bone window present in CT images. Bones typically lead to CT values between 350 and 3000 Hounsfield Units (HU) which serve as the described thresholds. The obtained set of voxels can be restricted to the area above the C5 vertebra which previously was obtained. Finally, we remove already predicted vertebras from the thresholded voxels which leaves us with an accurate mask for the skull. We furthermore exploit the behavior of the neural network predictions which have only been trained on parts of the anatomy and typically confuse structures that look similar in the CT images. Common systematic errors are to predict gonads as the eyeballs or colon as the nasal cavity. We exploit these systematic mistakes and remap the produced labels according to the location within the human body. By employing these simple rules we add $7$ additional labels.

\begin{figure}[t!]
\begin{tabular}{cc}
\begin{minipage}[t]{0.45\linewidth}
     \begin{algorithm}[H]
    \caption{- \textbf{Post-Processing}:  }
        \begin{algorithmic}
        \REQUIRE Volumetric Model Predictions
        \ENSURE Refined Volumetric Pseudo-Labels
        \STATE - \texttt{Left-Right Split}
        \STATE - \texttt{Rib Counting}
        \STATE - \texttt{Non-Largest Connected   Component \phantom{000} (CC) Supression}
        \STATE - \texttt{Area Restriction}
        \STATE - \texttt{Sex-based Consistency}
        \end{algorithmic}
    \label{Alg: Post-Processing}
    \end{algorithm} 
\end{minipage}
&  
\begin{minipage}[t]{0.45\linewidth}
     \begin{algorithm}[H]
    \caption{- \textbf{Sex-based Consistency}: \\
    Restrict reproductive anatomies based on sex (M/F).
    }
        \begin{algorithmic}
        \REQUIRE Reproductive Anatomy Predictions; Metadata
        \ENSURE Sex-Restricted Anatomy Predictions 
        \STATE - \texttt{Suppress M-reproductive anatomies\newline for F patients}
        \STATE - \texttt{Suppress F-reproductive anatomies\newline for M patients}
        \end{algorithmic}
        \label{Alg: Sex-based consistency}
    \end{algorithm} 
\end{minipage}

\\

\begin{minipage}[t]{0.45\linewidth}
     \begin{algorithm}[H]
    \caption{- \textbf{Non-largest CC Supression}: \\
    Maintain largest CC of anatomies only occurring once.
    }
        \begin{algorithmic}
        \REQUIRE Anatomy supposed to occur once
        \ENSURE Largest CC for each anatomy
        \STATE - \texttt{Identify 3D-CCs for the anatomy}
        \STATE - \texttt{Count voxels of each CC}
        \STATE - \texttt{Remove non-largest CC} 
        \end{algorithmic}
        \label{Alg: Non largest CC supression}
    \end{algorithm} 
\end{minipage}
&

\begin{minipage}[t]{0.45\linewidth}
     \begin{algorithm}[H]
    \caption{- \textbf{Area Restriction}: \\
    Restrict anatomy predictions to associated body part (BP)
    }
        \begin{algorithmic}
        \REQUIRE Predictions; BP associations; anchor classes
        \ENSURE Anatomy constrained by BP
        \STATE - \texttt{Define BP 
        based on box \newline around anchor classes}
        \STATE - \texttt{Bind predictions to 
        associated
        BP}
        \end{algorithmic}
    \label{Alg: Area Restriction}
    \end{algorithm} 
\end{minipage}
\\
 \begin{minipage}[t]{0.45\linewidth}
     \begin{algorithm}[H]
    \caption{- \textbf{Rib Counting}: \\
    Combat confusion between ribs.
    }
        \begin{algorithmic}
        \REQUIRE Rib Predictions
        \ENSURE  24 largest CC sorted by height
        \STATE - \texttt{Merge rib predictions}
        \STATE - \texttt{Apply Left-Right Split}
        \STATE - \texttt{Extract 24 largest 3D-CC} 
        \STATE - \texttt{Order CC by height of median points}
        
        \end{algorithmic}
    \label{Alg: Rib counting}
    \end{algorithm} 
\end{minipage}
&

\begin{minipage}[t]{0.45\linewidth}
    \begin{algorithm}[H]
        \caption{\textbf{- Left-Right Split:}\\
        Remap side-related labels based on the hyperplane. 
        }
        \begin{algorithmic}
        \REQUIRE Side-related labels; Sternum (S); Vertebrae (V)
        \ENSURE Side-related labels
        \STATE - \texttt{Fit hyperplane through V and S \newline centers to split left and right}
        \STATE - \texttt{Remap voxels through to their \newline center's position to the hyperplane}
        \end{algorithmic}
    \label{Alg: Left-Right-Splitting}
    \end{algorithm}
\end{minipage}
\end{tabular}

\caption*{To further improve the predicted label quality, we propose to use a post-processing procedure to improve the quality of the proposed labels. We apply a combination of a Left-Right-Split (assigning side-dependent labels to their correct side), Rib Counting (an ordering of the 24 human ribs), non-largest component suppression (for anatomies appearing with a single connected component like the brain), a form of area restriction (restricting anatomy labels to their rough body parts like thorax, abdomen, or head), and sex-based consistency (ensuring that patients are not presented with anatomies from a different sex).}
\label{fig:two-algorithms}
\end{figure}

\subsection*{Knowledge Aggregation}
In order to aggregate the predictions of the individual models, we define a common labeling scheme 
to which we map the obtained masks. Since some of these labels present multiple versions of the same anatomical structure, such as the class \textit{aorta} which is present in Total Segmentator, Amos, and SegThor it is necessary to combine these predictions. 
Unless stated otherwise, we merge the predictions of models trained on the different source datasets into a single mask which is the union of all individual masks. 
This procedure is simple and stable. It also helps in aggregating masks of the same anatomical structures which are only predicted within certain regions of the human body on which the respective models have been trained. An example of this behavior can be found in the mask for the class \textit{aorta} predicted by the SegThor~\cite{lambert2020segthor} model. While the aorta spans outside the thorax, this model only predicts it within the thorax region. Only by merging the mask for \textit{aorta} of this model with additional masks from other models leaves us with a full mask for the aorta spanning over the entire anatomy. Thus merging these labels combines the knowledge present in different parts of the human body into a single, unified anatomy which is the goal of this work. 

When integrating the different anatomical structures into the Atlas labeling scheme, we aggregate them according to their anatomical hierarchical level from course to fine starting from general tissues such as \textit{muscles} or \textit{fat}. On top, we gradually add the different organs and finally fine-grained vessel structures such as \textit{Pulmonary Arteries}. During the aggregation process, we employ basic anatomical knowledge to improve individual predictions on the fly. One of these operations is that we split the predicted voxels into left and right clusters for paired organs such as the hip bones, kidneys or adrenal glands and resolve conflicts. Furthermore, we restrict predictions based on previous labels or eliminate non-largest connected components if it is deemed appropriate. 


\subsubsection*{From Aggregated Predictions to a Unified Dataset}
\label{SS: Label Aggregation}
After integrating the labels into the common DAP Atlas CT volumes, it is an integrated dataset, but the different masks are still predictions of models which were trained on heterogeneous source datasets and thus generate heterogeneous masks. To unite these different, integrated masks into a single seamless dataset, we perform one iteration of self-training. The benefits of this procedure are four-fold: As previously mentioned, we bring the labels which originate from datasets of different resolutions into the common Atlas resolution leading to a truly seamless integration. A second reason to perform self-training is to eliminate non-systematic random noise. The network receives consistent feedback from consistent predictions, while noisy predictions are non-systematic. This observation is a well-known fact in image classification~\cite{liu2020early} which states that before the memorization of training data, networks tend to ignore noisy predictions and focus on consistent feedback. We make sure to not overfit the network on the dataset by closely monitoring training and validation losses. A third reason, to perform self-training is to distill the fragmented knowledge into a single model capable to predict the entire anatomy, this massively decreases the necessary time to predict the anatomy, since it reduces the inference time from $n$ expert models to a single model. Finally, self-training hampers the exact reconstruction of private data from expert models which were directly trained on private source datasets. 

We generate the first version of the DAP Atlas dataset by applying the obtained unified anatomical model on the selected Atlas target volumes. While the overall label quality is good, we notice certain patterns which were repeatedly done wrong and with which the networks seemed to struggle. These systematic exceptions are the confusion of voxels that belong to paired structures such as the left and right kidney or adjacent vertebrae. Further, we observe implausible predictions of structures within body regions that are not possible, e.g. colon being predicted outside the abdomen. Finally, we observe structures belonging to the reproductive system to be predicted for the wrong sex. 
These errors are relatively easy to correct by once again applying anatomical rules. To address these, we propose Algorithm~\ref{Alg: Post-Processing}. 
We furthermore use two sets of rules to filter out implausible prediction: During Algorithm ~\ref{Alg: Rib counting}: We exclude predictions leading to different orderings induced by median points and minimum points of the ribs. Additionally, we examine the normal vector of the hyperplane during Algorithm~\ref{Alg: Left-Right-Splitting} and exclude predictions leading to hyperplanes that deviate too much from the axial directions. This reduces the number of images in the Atlas dataset from $566$ to $533$ CTs.

After applying Algorithm~\ref{Alg: Post-Processing} to the raw labels, we receive the final version of the dataset, which is rated as very impressive by a consulted radiologist. We describe the extensive validation procedure of the dataset in Section \textit{Technical Validation}. 

While the dataset is convincing, we acknowledge that the performance of the developed anatomical model is dependent on Algorithm~\ref{Alg: Post-Processing} which is undesirable as it requires the availability of anchor predictions which may not always be available for arbitrary CTs. As an additional contribution besides the dataset, we develop a more robust, model based on the available Atlas Knowledge which is more suitable for a clinical environment in which the model has to process arbitrary CT Volumes. We will refer to the previous model used to generate the dataset as the Atlas dataset model (V1) and the novel model as the Atlas prediction model (V2).

\subsubsection*{Developing a Prediction Model from the Atlas Dataset}
The goal of the Atlas prediction model is to eliminate the need for post-processing which is impractical within a clinical setting in which the model should be able to deliver convincing results on arbitrary CT volumes. When examining the different steps of Algorithm~\ref{Alg: Post-Processing}, we notice two steps that are easy to address algorithmically: sex-based consistency and non-largest connected component suppression as defined in Algorithms~\ref{Alg: Sex-based consistency} and~\ref{Alg: Non largest CC supression} respectively, as these methods simply suppress predictions and do not rely on other anchor predictions such as Algorithm~\ref{Alg: Left-Right-Splitting}. We thus aim to develop a training procedure that eliminates the need for left-right splitting (Algorithm~\ref{Alg: Left-Right-Splitting}), area-restrictions (Algorithm~\ref{Alg: Area Restriction}), and rib counting (Algorithm~\ref{Alg: Rib counting}).

\begin{figure}
    \centering
    \includegraphics[trim=0 100 000 000, clip, width=0.8\textwidth]{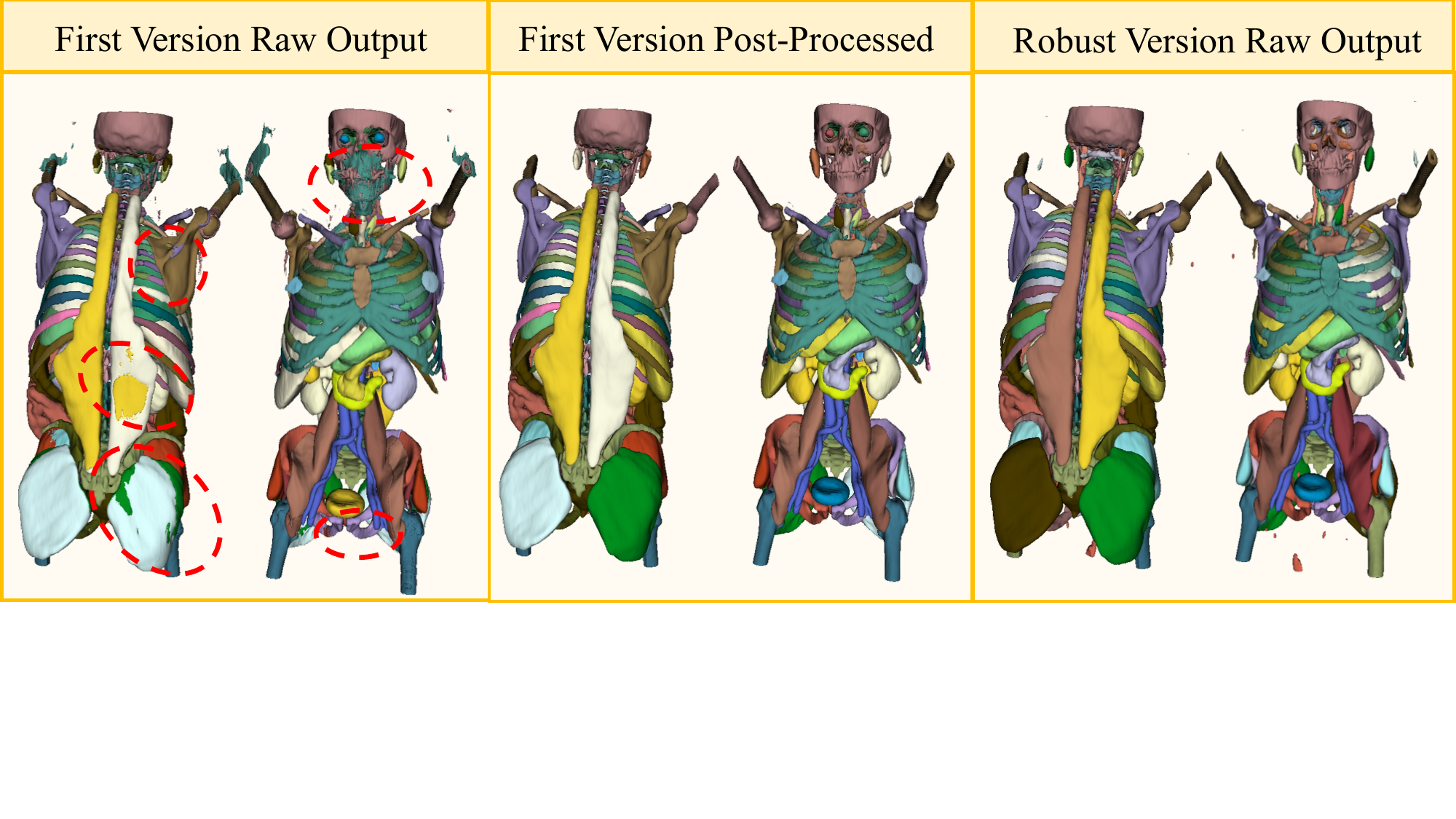}
    \caption{Comparison of the raw output of the standard unified DAP Atlas dataset model, the post-processed volume using Algorithm~\ref{Alg: Post-Processing} and the raw output of the more robust Atlas prediction model. In red we mark problematic regions in the raw labels obtained by the first version of the model. The post-processed volumes and the raw model volumes are alike. The color encoding of the different structures between the Standard Version and the Robust Version is arbitrary and does not have a meaning.}
    \label{fig:V1_V2_comparison}
\end{figure}

To tackle these challenges we develop a custom training strategy for the Atlas prediction model. First, we apply Algorithm ~\ref{Alg: Post-Processing} during the aggregation phase of the individual expert models to maximize the agreement with the desired output which has been approved by experts. Next, we observe that due to the large number of classes, the standard nnU-Net~\cite{isensee2021nnu} learning rate schedule is suboptimal as it closely follows a linear learning rate schedule allocating approximately the same number of epochs for small and large learning rates. We find that the proposed task is more difficult than most standard segmentation tasks and thus increase the number of training epochs from $1000$ to $5000$. Finally, we fine-tune the network for another $1000$ epochs with a fixed learning rate of $0.001$ and without the standard mirror augmentation. This allows the network to focus on the improvements on smaller structures and helps to mitigate the right-left and rib confusion. We show a comparison of the raw output of the Atlas dataset model, the post-processed volume, and the raw output of the Atlas prediction model in Fig.~\ref{fig:V1_V2_comparison}. As it can be seen, the output of the robust model has a large agreement with the post-processed predictions of the first model without relying on Algorithm~\ref{Alg: Post-Processing}. We analyze this behavior and find that the vast majority of predicted structures have an agreement of more than $90\%$ IoU between the post-processed V1 Model and the raw V2 predictions.

Besides the DAP Atlas dataset, we also release the robust segmentation model which can be used to perform inference without post-processing. It furthermore tends to perform better for out-of-distribution tasks which are common within a clinical setting. We examine this behavior in Section \textit{Technical Validation}.

%% file: tabel_labels.tex
\begin{figure}[t]
    \centering
    \includegraphics[width=0.8\linewidth,height=0.6\linewidth]{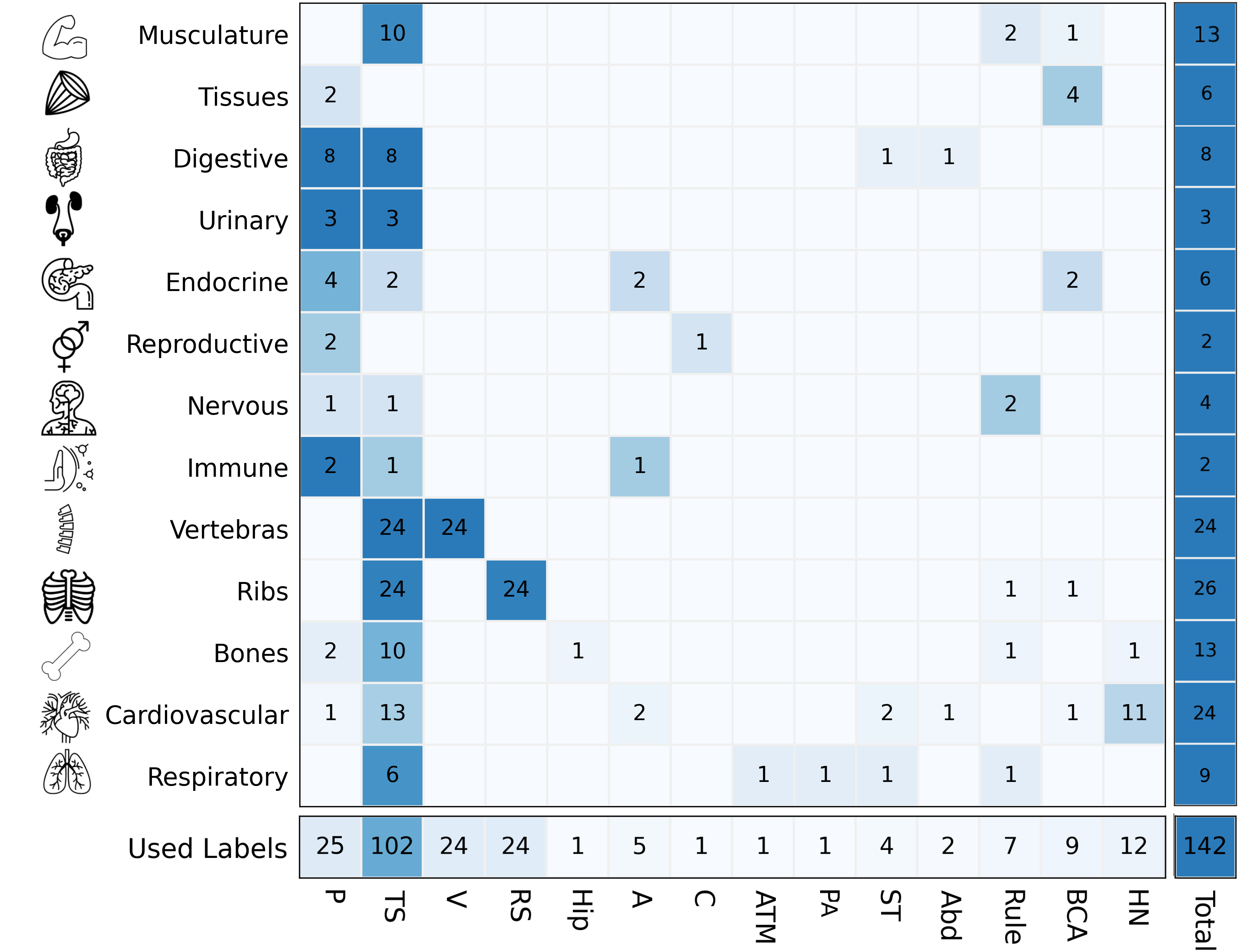}
    \caption{Overview of the different source datasets from which the DAP Atlas dataset is derived. Within this table, we cluster the individual labels according to their anatomical system. We show a full table with individual labels in the supplementary material. On the bottom row, we show the number of labels in the DAP Atlas dataset which are influenced by the respective source dataset. The sum of columns cannot be calculated, because predictions of the same label are fused. The abbreviations for the datasets are as follows: \textbf{P}:Pediatric~\cite{jordan2022pediatric}, \textbf{TS}: Total Segmentator~\cite{wasserthal2022totalsegmentator}, \textbf{V}: Verse~\cite{sekuboyina2021verse}, \textbf{RS}:RibSeg~\cite{yang2021ribseg}, \textbf{Hip}: Pelvis CT~\cite{liu2021deep}, \textbf{A}:Amos~\cite{ji2022amos}, \textbf{C}:MAL Cervix~\cite{landman2015miccai}, \textbf{ATM}: ATM Airway Tree Modelling~\cite{zhang2023multi}, \textbf{PARSE}: Pulmonary Artery Segmentation Challenge~\cite{kuanquan_wang_2022_6361906}, \textbf{ST}:SegThor~\cite{lambert2020segthor}, \textbf{Abd}: CT50 Abdomen~\cite{ma2021abdomenct}, \textbf{Rule}: Rule based derived label, \textbf{BCA}: Body composition analysis model~\cite{koitka2021fully}, \textbf{HN}: Private Head-neck dataset. 
    }
    \label{Tab:Label_Usage}
\end{figure}

%% file: Latex/Data_Records.tex
\section*{Data Records}
\label{S: Data Record}

We make use of the recently published AutoPET dataset~\cite{gatidis2022whole} which can be accessed on The Cancer Imaging Archive (TCIA) under its collection name “FDG-PET-CT-Lesions” to download the raw PET/CT data. We publish the segmentation masks representing the DAP Atlas dataset at \href{https://github.com/alexanderjaus/AtlasDataset}{github}. In our DAP Atlas dataset, we have retained the AutoPET naming convention to ensure that the masks can be easily matched with their corresponding original CT volume.
\vspace{5mm}

\dirtree{%
.1 DAP Atlas Anatomical Labels.
.2 AutoPET\_0011f3deaf\_10445.nii.gz.
.2 AutoPET\_01140d52d8\_56839.nii.gz.
.2 AutoPET\_0143bab87a\_33529.nii.gz.
.2 \dots .
}
\vspace{5mm}
 The given name consists of the subject ID followed by the last $5$ digits of the Study UID which allows a unique matching of the segmentation masks to the AutoPET CTs. 

%% file: Latex/Technical_Validation.tex
\section*{Technical Validation}
\label{S: Technical Validation}

As previously discussed, we propose the DAP Atlas dataset as a knowledge aggregation dataset from multiple fragmented source datasets, which are impractical to train neural networks on, as they only offer partial supervision for the presented anatomical structure and label everything else as background. As the DAP Atlas dataset consists of many volumes and is rich in labels, it is nearly impossible to have experts check every voxel for correctness. As previously mentioned, other datasets containing few annotations can still use manual label checking and correction. The Airway Tree Modeling dataset~\cite{zhang2023multi}, which is comparable in size, provides annotations for a single label. With the previously discussed arguments in Section~\textit{Background and Summary}, the creation time was $80$ radiologist days. This example demonstrates that even checking and correcting for the DAP atlas dataset on a voxel level by humans is nearly impossible.
The TotalSegmentator dataset~\cite{wasserthal2022totalsegmentator} proposes to use 2D renderings of 3D organs to improve the required time to check for correctness. This type of shape check gains in speed, but it also does not guarantee the correctness of each voxel. 

\subsection*{Evaluation Setup:}
To tackle the aforementioned problem of evaluation, we propose a hybrid approach combining human experts, anatomical plausibility, and usefulness for the Deep Learning community. 
\begin{itemize}
    \item \textbf{Deep Learning Applicability:} We verify the usefulness of our dataset for the development of Deep Learning algorithms by taking our anatomical segmentation models which were trained on DAP Atlas and perform inference on the BTCV~\cite{landman2015miccai} abdomen dataset. This dataset has not been used for the dataset construction and provides an unbiased performance check. We compare the performance of the Atlas dataset model and the Atlas prediction model. 
    \item \textbf{Expert Checks:} We sample $25$ volumes from the DAP Atlas dataset and let an expert radiologist evaluate them. We ask the radiologist to report on overall label quality and provide insights on potential use.
    \item \textbf{Anatomical Insights of DAP Atlas:} To verify the general, global anatomical plausibility of the dataset, we use the labels of the DAP Atlas dataset to calculate the volumes and mean intensities as characteristic descriptors of the different anatomical structures. We plot these descriptors against the age and gender of the patients and identify if they follow characteristic medical curves. 
    Further, we compare the volume distributions of Atlas organs with several source datasets to investigate the deviation of the different volume distributions. Finally, we investigate which anatomical structures in the Atlas dataset are most affected by which type of cancer.
    
\end{itemize}

By combining these three approaches we combine the thoroughness of local, voxel-wise checks with the scalability of global overall checks and make sure that the dataset introduces merit to the Deep Learning community. 

\subsection*{Results:}
\subsubsection*{Deep Learning Applicability}

\begin{figure}[t]
    \centering
    \includegraphics[trim=0 390 420 0, clip, width=0.95\textwidth]{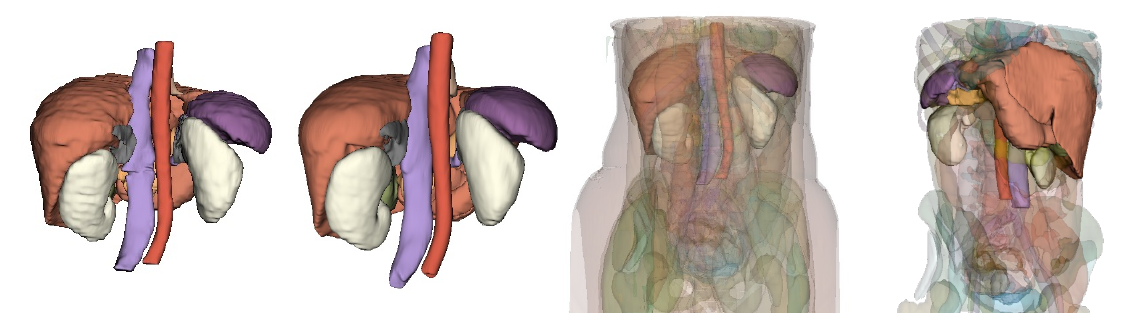}
    \caption{Qualitative assessment of the BTCV~\cite{landman2015miccai} out-of-distribution performance of Atlas prediction model. Since we do not have access to the labels of the test dataset, we perform inference on the training dataset which our model has not seen either. On the very left we show the ground truth of an example within the training dataset next to the predictions obtained by our model. On the right, we show all of the predictions which can be obtained by our model and highlight the $13$ BTCV labels. We make two observations. The $13$ BTCV organs are meticulously addressed in our approach, aligning closely with the ground truth. Furthermore, the BTCV organs are only a fraction of the structures that our model is able to segment. Our model achieves a comprehensive, dense anatomical segmentation of the human body.}
    \label{fig:BTCV Inference}
\end{figure}

Regarding the usefulness of the provided dataset for Deep Learning models, we use our unified anatomical models to predict the Atlas anatomy onto the BTCV~\cite{landman2015miccai} abdomen dataset. This dataset has not been used to construct the DAP Atlas dataset and provides an unbiased performance check. We emphasize, that we do not fine-tune the models using the BTCV training data, but perform inference on the BTCV testset without post-processing. 
We report the performance measures of the Atlas dataset model and the Atlas prediction model in Table~\ref{tab:btcv}. 

Our Atlas dataset model achieves an average dice score of about $81\%$. 
The Atlas prediction model (V2), developed through iterative training and post-processing cycles, achieves an $85\%$ dice score, an improvement of $\sim 4\%$, demonstrating its increased robustness due to the adapted training schedule. $85\%$ dice is on par with state-of-the-art medical segmentation models such as UNETR~\cite{hatamizadeh2022unetr} which are trained in a standard supervised fashion on the training dataset. 
We see that the V2 model shows significant improvements in abdominal structures, i.e. $81\%$ vs $85\%$ for the \textit{vena cava inferior (IVC)} or $75\%$ vs $83\%$ for the pancreas and in particular smaller structures such as \textit{Adrenal Glands}. 
But we also notice small decreases in performance for \textit{Left and Right Kidneys}. 
Regarding the Mean Surface Distance performance, we see an overall improvement of the Atlas V2 model compared to the Atlas dataset model (V1), however, the individual organs improvements do not follow a clear pattern. 
In summary, we find that the DAP Atlas dataset allows for the creation of high-quality anatomical models and our proposed adapted training strategy for the development of a more robust Atlas V2 model as described previously seems to deliver the expected results.

\input{btcv_perf}

A qualitative assessment of the label quality and the provided labels beyond the required labels is shown in Fig.~\ref{fig:BTCV Inference}. Our model predictions exhibit a remarkable level of alignment when compared to the ground truth. The only noticeable differences are the slightly smoother surface compared to the ground truth and minor shape differences in the liver.
Beyond that, we also show how the BTCV organs are a well-integrated fraction of the anatomical structures of the human body which our model is able to segment.
\subsubsection*{Expert Checks:}
We include a human expert in the quality check pipeline. 
To gather human feedback on the DAP Atlas dataset, we randomly sampled $25$ volumes and let an expert radiologist evaluate the quality and discuss shortcomings and applications. The following section discusses the feedback we received and displays the strengths and weaknesses of the DAP Atlas dataset.
\\The feedback that we received was mostly positive:

\begin{quote}
\textit{
    Overall, for whole-body segmentation of a normal patient, it's very impressive. [...] It was also good on some patients pointing out a small hiatal hernia. Otherwise, I think it is more useful for medical students and internal medicine doctors who may not be as familiar with anatomy on CT.}
\end{quote}

Besides this general feedback, we gained some insights into the structural mechanics of the dataset, which we summarize in the following. The expert noted that some structures seem to not always be homogeneously segmented, naming predominantly the spinal canal. Further, for tree-like structures such as the pulmonary artery, the fine-grained branch endings lose detail and become under-segmented. Lastly, it was noted, that the borders of abutting abdominal organs are at times offset and differ from the expert's estimation.



\subsubsection*{Anatomical Insights of DAP Atlas:}
\begin{figure}
    \centering
    \includegraphics[trim=120 100 140 120, clip, width=0.95\textwidth]{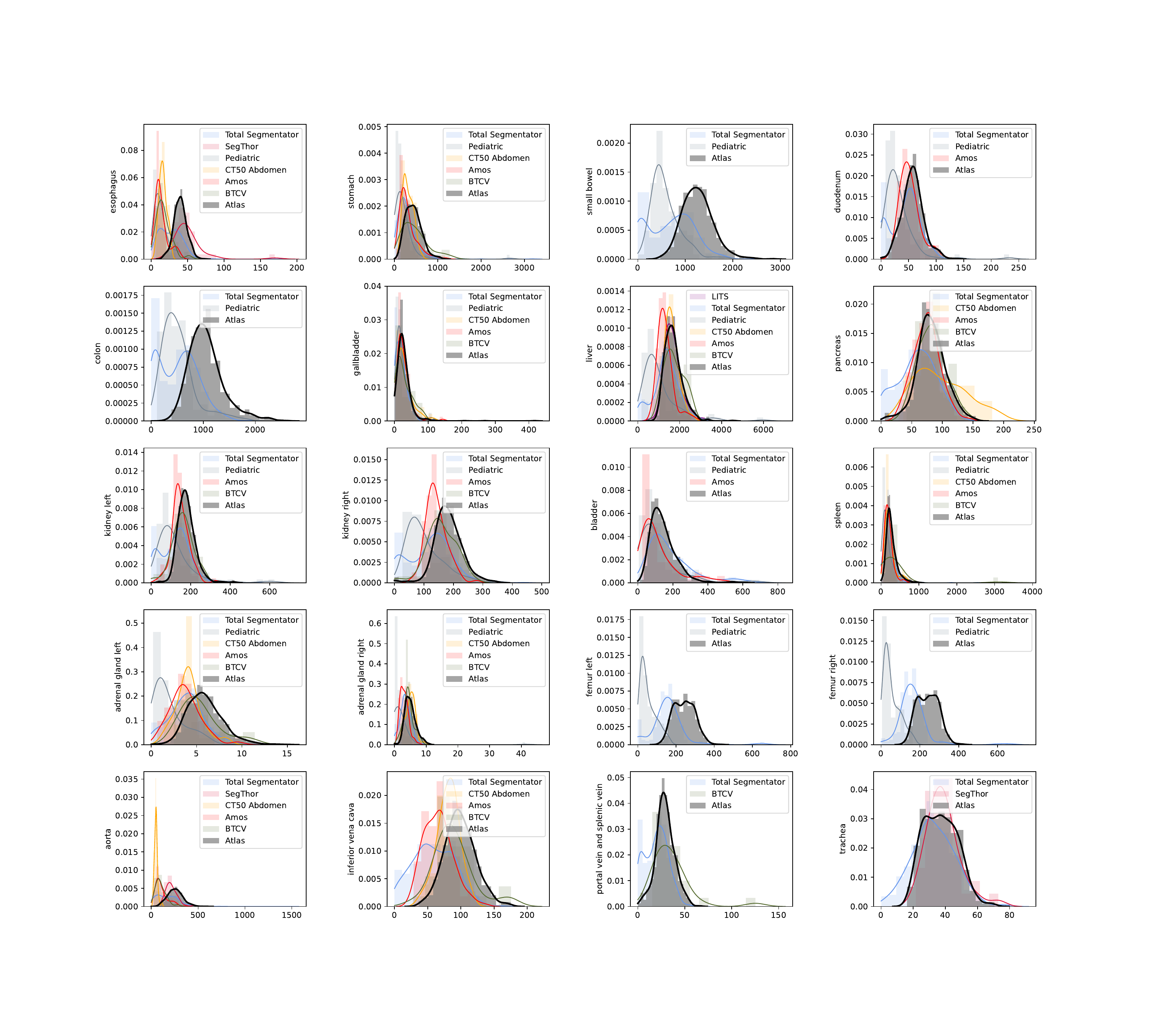}
    \includegraphics[width=0.8\textwidth]{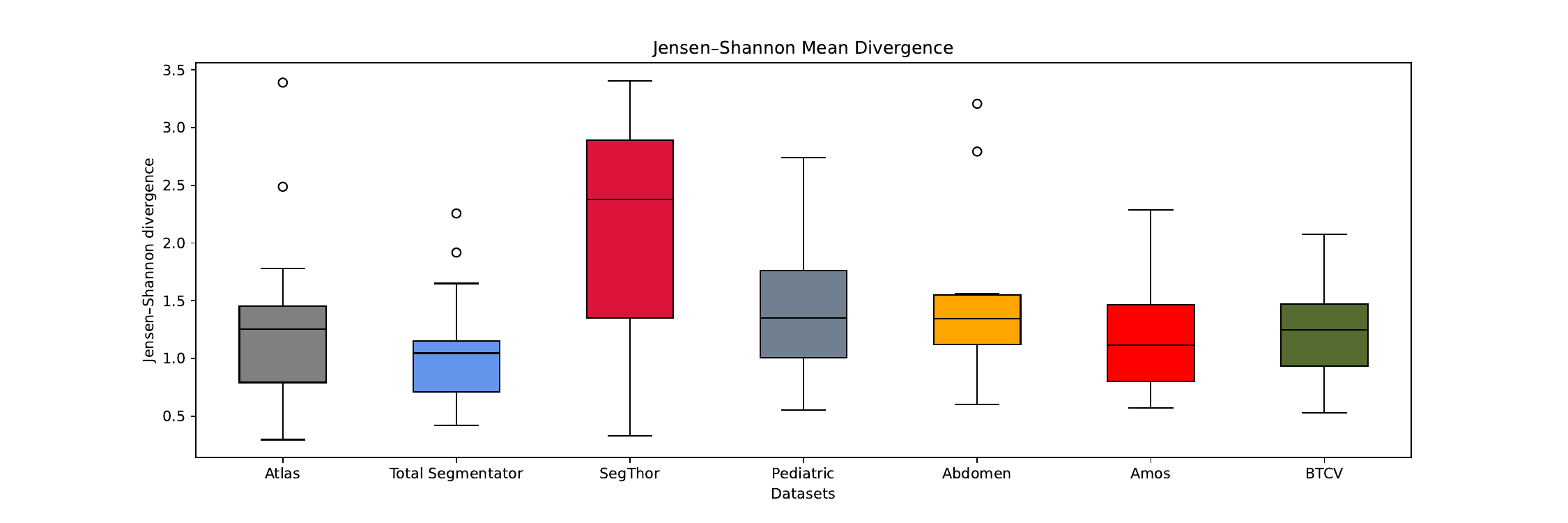}
    \caption{Comparisons of the distribution of volumes in milliliter for exemplary anatomical structures for several source datasets and DAP Atlas. We analyze the distribution of the average Jensen-Shannon Divergence (JSD) among the different distributions calculated for each anatomical structure in the box-plot below. The lower the average JSD, the better the agreement with other datasets.}
    \label{fig:All_distplots}
\end{figure}

As a first global check, we compare the volume distributions of our proposed DAP Atlas dataset against other datasets which were annotated by experts. The different volume distributions are shown in Fig.~\ref{fig:All_distplots}. We find that our proposed dataset is placed well within the volume distributions of other expert datasets in both distribution shape and distribution support. 

The selection of anatomical structures for which we show the distribution plots is based on the criteria that at least two additional datasets next to the proposed DAP Atlas dataset contain the structure. By observing the distributions we find, that the volume distributions for the same organs in different datasets do vary by small amounts, but the general shape of the distributions are very similar among the datasets. Small differences in the distributions of organ volumes across the datasets are quite plausible and may stem from limited samples, different annotation schemes, or the selection criteria of the patients. For instance, the distributions of organs in the Pediatric dataset tend to be shifted to the left, which can be easily explained since the dataset focuses on patients below the age of $18$ years. Larger variations and in particular distributions deviating towards the origin can stem from CT images only covering parts of organs which is common in the Total Segmentator dataset. When comparing the DAP Atlas dataset to the family of organ distributions, we find it to be well-integrated regarding its distribution support and shape.  

To analytically confirm this, we calculate the Jensen–Shannon Divergence (JSD) as a symmetric, finite measure to calculate the deviations of distributions. For each of the analyzed anatomical structures, we calculate the JSD of a dataset's volume distribution to all other volume distributions within the same structure. We average these values to receive the distribution's average distance to all other distributions. The greater the average value, the more distant the volume distribution is from its peers. Finally, we draw a box-plot to compare the distribution of average JSD distances per dataset. We find that the DAP Atlas dataset is well placed among the other datasets with distributional agreements very similar to those of voxel-wise expert annotated datasets.

\begin{figure}
    \centering
    \includegraphics[trim=0 0 0 0, clip, width=0.95\textwidth]{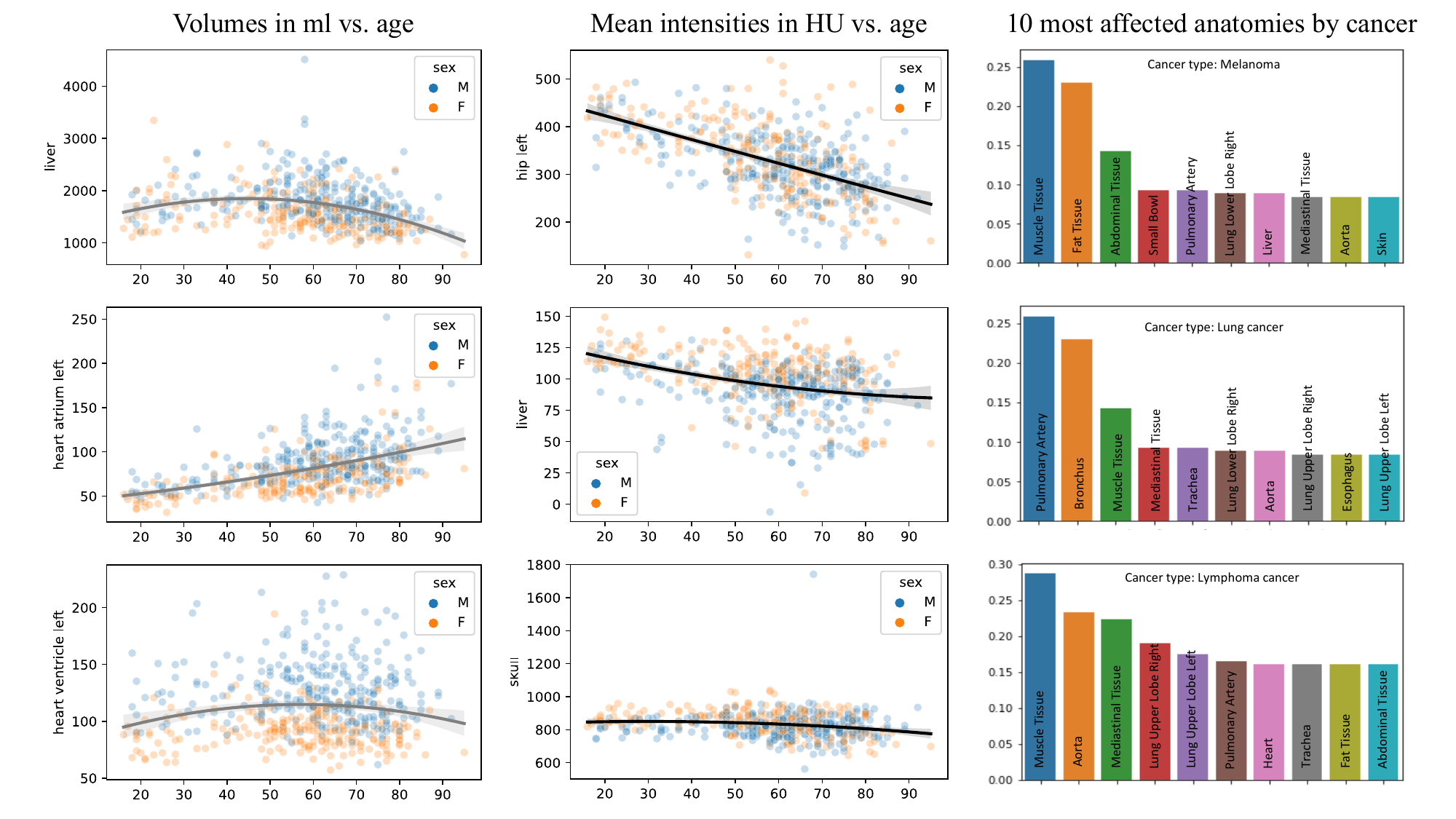}
    \caption{On the left, we show the volume in milliliters of characteristic organs of the DAP Atlas dataset plotted against the age of the patients. While the volume of the liver follows a downward-facing parabola with increasing and decreasing characteristics, the heart atrium tends to only increase with progressing age. Both behave akin to established knowledge~\cite{keller2021right}. 
    In the middle column, we show the calculated mean intensities of the respective structure in the CT volume indicated by our Atlas labels. 
    In the right column, we examine which known organ structures in the Atlas dataset are most affected by which type of cancer. 
    }
    \label{fig:Age_distplots}
\end{figure}

As a second global check, we calculate the volume and the mean intensity of each of the anatomical structures in the DAP Atlas dataset and plot them against the age of the patients in Fig.~\ref{fig:Age_distplots}. 

On the left, we show the volume in milliliters of characteristic organs of the DAP Atlas dataset plotted against the age of the patients. We observe that the volumes of organs for female patients tend to be smaller compared to the organs of men. Further, we fit a quadratic model to explain organ volumes as a function of age and find plausible medical relationships. Whereas the volume of the liver follows a downward-facing parabola with increasing and decreasing characteristics, the hearth atrium tends to only increase with progressing age. Both of these behaviors are well-known medical facts~\cite{keller2021right}, confirming the anatomical plausibility of the dataset.

In the middle column of Fig.~\ref{fig:Age_distplots}, we show the calculated mean intensities of the respective structure in the CT volume indicated by our
atlas labels. We choose three exciting examples and observe very plausible curves when examining the relationship between the mean-intensities of the atlas labels and the age of the patients. For instance, the relationship between hip bone density and age almost linearly decreases with increasing age due to osteoporosis. Organ tissues also tend to become less dense with increasing age. Finally, an interesting observation can be derived from the last plot of the middle column in which we examine the reported outlier. The reason for this extreme behavior is dental implants pushing up the mean intensity of the skull.


In the right column, we show an example of a potential future use case in which the Atlas dataset may serve as a cornerstone in the joint investigation of the entire anatomy and pathologies. We calculate which of the known structures in the Atlas dataset are most affected by which type of cancer. For each patient in the Atlas dataset, we examine which anatomical structures are affected by cancer. When determining if an anatomical structure has been affected by cancer, we consider it to be cancerous if there is at least one voxel labeled as cancerous tissue. During this analysis, we do not distinguish between metastasis and primary cancer cells. Finally, we normalize by the total number of patients with the respective disease to obtain how likely it is that an anatomical structure is affected given the respective diagnosis.


\input{full_list_of_labels}

%% file: btcv_perf.tex
\setlength{\tabcolsep}{2pt}
\begin{table}[h]
\centering
\footnotesize
\begin{tabular}{lccccccccccccccc}
\toprule

& Spleen & RKidney & LKidney & Gallbladder & Esophagus & Liver & Stomach & Aorta & IVC & PV\&SV & Pancreas & RAdrenal & LAdrenal & Total\\
\midrule
\multicolumn{15}{c}{Dice Scores}\\
\midrule
Atlas (V1) & 0.96 & 0.91 & 0.94 & 0.62 & 0.72 & 0.97 & 0.84 & 0.91 & 0.81 & 0.76 & 0.75 & 0.64 & 0.59 &  
0.81\\
Atlas (V2) & 0.96 & 0.86 & 0.92 & 0.72 & 0.80 & 0.96 & 0.86 & 0.92 & 0.85 & 0.77  & 0.83 & 0.75 & 0.74 & 
0.85\\
\midrule
\multicolumn{15}{c}{Mean Surface Distance}\\
\midrule
Atlas (V1) & 1.14 & 2.21 & 0.83 & - & 1.68 & 0.79 &  3.26  & 1.51 & 2.03 & 1.48 & 2.53 & 1.30 & 1.67 & 2.03\\
Atlas (V2) & 0.65 & 4.28 & 1.70 & - & 1.38 & 1.21 &  2.50 & 1.29 & 1.87 & 1.77 & 1.51 & 0.80 & 0.90 & 1.85\\
\bottomrule
\hline
\end{tabular}
\caption{Class-wise Dice and Mean Surface Distance performance on the BTCV dataset for the standard Atlas model and the robust Atlas model. Both predictions have not been post-processed. 
It can clearly be seen that the robust Atlas V2 model benefited from the adapted training procedure.}
\label{tab:btcv}

\end{table}

%% file: full_list_of_labels.tex
\begin{table}[htbp]
  \centering
  \begin{tabular}{|c|c|c|c|c|c|c|c|}
    \hline
    \textbf{ID} & \textbf{Label} & \textbf{ID} & \textbf{Label} & \textbf{ID} & \textbf{Label} & \textbf{ID} & \textbf{Label} \\
    \hline
    \hline
    0                            & Background                      & 38                           & Iliopsoas Left                  & 75                           & Costa 5 Left                    & 112                          & Iliac Artery Left               \\
1                            & Unknown Tissue                & 39                           & Iliopsoas Right                 & 76                           & Costa 5 Right                   & 113                          & Iliac Artery Right              \\
2                            & Muscles                         & 40                           & Autochthon Left                 & 77                           & Costa 6 Left                    & 114                          & Aorta                           \\
3                            & Fat                             & 41                           & Autochthon Right                & 78                           & Costa 6 Right                   & 115                          & Iliac Vena Left                 \\
4                            & Abdominal Tissue                & 42                           & Skin                            & 79                           & Costa 7 Left                    & 116                          & Iliac Vena Right                \\
5                            & Mediastinal Tissue              & 43                           & Vertebrae C1                    & 80                           & Costa 7 Right                   & 117                          & Inferior Vena Cava              \\
6                            & Esophagus                       & 44                           & Vertebrae C2                    & 81                           & Costa 8 Left                    & 118                          & Portal Vein and Splenic Vein    \\
7                            & Stomach                         & 45                           & Vertebrae C3                    & 82                           & Costa 8 Right                   & 119                          & Celiac Trunk                    \\
8                            & Small Bowel                     & 46                           & Vertebrae C4                    & 83                           & Costa 9 Left                    & 120                          & Lung Lower Lobe Left            \\
9                            & Duodenum                        & 47                           & Vertebrae C5                    & 84                           & Costa 9 Right                   & 121                          & Lung Upper Lobe Left            \\
10                           & Colon                           & 48                           & Vertebrae C6                    & 85                           & Costa 10 Left                   & 122                          & Lung Lower Lobe Right           \\
12                           & Gallbladder                     & 49                           & Vertebrae C7                    & 86                           & Costa 10 Right                  & 123                          & Lung Middle Lobe Right          \\
13                           & Liver                           & 50                           & Vertebrae T1                    & 87                           & Costa 11 Left                   & 124                          & Lung Upper Lobe Right           \\
14                           & Pancreas                        & 51                           & Vertebrae T2                    & 88                           & Costa 11 Right                  & 125                          & Bronchus                        \\
15                           & Kidney Left                     & 52                           & Vertebrae T3                    & 89                           & Costa 12 Left                   & 126                          & Trachea                         \\
16                           & Kidney Right                    & 53                           & Vertebrae T4                    & 90                           & Costa 12 Right                  & 127                          & Pulmonary Artery                \\
17                           & Bladder                         & 54                           & Vertebrae T5                    & 91                           & Rib Cartilage                   & 128                          & Cheek Left                      \\
18                           & Gonads                          & 55                           & Vertebrae T6                    & 92                           & Sternum Corpus                  & 129                          & Cheek Right                     \\
19                           & Prostate                        & 56                           & Vertebrae T7                    & 93                           & Clavicula Left                  & 130                          & Eyeball Left                    \\
20                           & Uterocervix                     & 57                           & Vertebrae T8                    & 94                           & Clavicula Right                 & 131                          & Eyeball Right                   \\
21                           & Uterus                          & 58                           & Vertebrae T9                    & 95                           & Scapula Left                    & 132                          & Nasal Cavity                    \\
22                           & Breast Left                     & 59                           & Vertebrae T10                   & 96                           & Scapula Right                   & 133                          & Artery Common Carotid Right     \\
23                           & Breast Right                    & 60                           & Vertebrae T11                   & 97                           & Humerus Left                    & 134                          & Artery Common Carotid Left      \\
24                           & Spinal Canal                    & 61                           & Vertebrae T12                   & 98                           & Humerus Right                   & 135                          & Sternum Manubrium               \\
25                           & Brain                           & 62                           & Vertebrae L1                    & 99                           & Skull                           & 136                          & Artery Internal Carotid Right   \\
26                           & Spleen                          & 63                           & Vertebrae L2                    & 100                          & Hip Left                        & 137                          & Artery Internal Carotid Left    \\
27                           & Adrenal Gland Left              & 64                           & Vertebrae L3                    & 101                          & Hip Right                       & 138                          & Internal Jugular Vein Right                      \\
28                           & Adrenal Gland Right             & 65                           & Vertebrae L4                    & 102                          & Sacrum                          & 139                          & Internal Jugular Vein Left                       \\
29                           & Thyroid Left                    & 66                           & Vertebrae L5                    & 103                          & Femur Left                      & 140                          & Artery Brachiocephalic          \\
30                           & Thyroid Right                   & 67                           & Costa 1 Left                    & 104                          & Femur Right                     & 141                          & Vein Brachiocephalic Right     \\
31                           & Thymus                          & 68                           & Costa 1 Right                   & 105                          & Heart                           & 142                          & Vein Brachiocephalic Left      \\
32                           & Gluteus Maximus Left            & 69                           & Costa 2 Left                    & 106                          & Heart Atrium Left               & 143                          & Artery Subclavian Right        \\
33                           & Gluteus Maximus Right           & 70                           & Costa 2 Right                   & 107                          & Heart Tissue                    & 144                          & Artery Subclavian Left         \\
34                           & Gluteus Medius Left             & 71                           & Costa 3 Left                    & 108                          & Heart Atrium Right              &                              &                                 \\
35                           & Gluteus Medius Right            & 72                           & Costa 3 Right                   & 109                          & Heart Myocardium                &                              &                                 \\
36                           & Gluteus Minimus Left            & 73                           & Costa 4 Left                    & 110                          & Heart Ventricle Left            &                              &                                 \\
37                           & Gluteus Minimus Right           & 74                           & Costa 4 Right                   & 111                          & Heart Ventricle Right           &                              &                                 \\ \hline
\end{tabular}
\label{Tab: All Labels}
\caption{Full list of available labels in the DAP Atlas dataset. Besides the self-explanatory tissues, we include a class \textit{Unknown Tissue} indicating tissue that likely still needs to be annotated. It typically contains tissue structures that have not been annotated explicitly but were obtained by morphological operations. We still include this class as it has the potential to be useful for future work.}
\end{table}

%% file: Latex/Discussion.tex
\section*{Discussion and Conclusion}
\label{S: Discussion}
In this work, we generated the full-body DAP Atlas dataset through the aggregation of fragmented knowledge in combination with self-training, guided by medicine-derived rules.  
The dataset enables the training of holistic anatomy segmentation models in CT, which we evaluated through performances in transfer learning, expert inspection, and comparability to established knowledge about the human body.

We have shown that the proposed DAP Atlas dataset carries valuable, accurate anatomical knowledge which can be used to train neural networks capable to perform high-quality segmentation on previously unseen data. The proposed robust training procedure of the Atlas prediction model outperforms the standard nnU-Net training in out-of-distribution tasks confirming its usefulness for general prediction tasks. 

We believe in the future of only limited expert supervision as it will no longer be possible to create large-scale datasets matching those of natural images in the medical field due to the required expertise. Thus, any semi-automatic validation approach comes with the downside that the correctness of each individual voxel cannot be guaranteed. 
However, we want to emphasize, that non-systematic noise behavior can be handled for instance via the regularization approach described in Section \textit{Methods}.

Based on the global checks and the Atlas Model's strong performance on BTCV, we can confidently say that the expert's favorable opinion likely applies to the other CT volumes that have not undergone voxel-wise analysis. 






Our proposed expert-free dataset generation approach was used to construct our full-body DAP Atlas dataset which is a large, full-body CT dataset with dense annotations for most of the important anatomical structures. Our methodology involves consolidating existing but scattered information found in partially annotated datasets, using self-training methods guided by medical principles. 

We point out, that the future of large-scale medical datasets will most likely be expert free at least in the sense of voxel-wise alignment checks due to the inconvenient combination of limited experts in the medical field and required expertise to guarantee voxel-wise correctness. Thus, we propose a hybrid approach to evaluate the proposed DAP Atlas dataset, partially combining voxel-wise alignment checks on randomly sampled subsets and automated checks. We argue that due to the global overall anatomical consistency and the positive feedback on a random sample, the overall quality of the dataset is convincing which is further supported by the already mentioned impressive performance on the BTCV benchmark without leveraging its training dataset. 


To foster progress on dense anatomical predictions in the medical community, both, the dataset and the models will be made publicly available.

Overall, we believe that this work is a first step towards automated large-scale dataset generation and validation with the usage of anatomical knowledge to overcome the limited expert problem within the medical field.

%% file: main.bbl
\begin{thebibliography}{10}
\urlstyle{rm}
\expandafter\ifx\csname url\endcsname\relax
  \def\url#1{\texttt{#1}}\fi
\expandafter\ifx\csname urlprefix\endcsname\relax\def\urlprefix{URL }\fi
\expandafter\ifx\csname doiprefix\endcsname\relax\def\doiprefix{DOI: }\fi
\providecommand{\bibinfo}[2]{#2}
\providecommand{\eprint}[2][]{\url{#2}}

\bibitem{ronneberger2015u}
\bibinfo{author}{Ronneberger, O.}, \bibinfo{author}{Fischer, P.} \&
  \bibinfo{author}{Brox, T.}
\newblock \bibinfo{title}{U-net: Convolutional networks for biomedical image
  segmentation}.
\newblock In \emph{\bibinfo{booktitle}{Medical Image Computing and
  Computer-Assisted Intervention--MICCAI 2015: 18th International Conference,
  Munich, Germany, October 5-9, 2015, Proceedings, Part III 18}},
  \bibinfo{pages}{234--241} (\bibinfo{organization}{Springer},
  \bibinfo{year}{2015}).

\bibitem{cciccek20163d}
\bibinfo{author}{{\c{C}}i{\c{c}}ek, {\"O}.}, \bibinfo{author}{Abdulkadir, A.},
  \bibinfo{author}{Lienkamp, S.~S.}, \bibinfo{author}{Brox, T.} \&
  \bibinfo{author}{Ronneberger, O.}
\newblock \bibinfo{title}{3d u-net: learning dense volumetric segmentation from
  sparse annotation}.
\newblock In \emph{\bibinfo{booktitle}{Medical Image Computing and
  Computer-Assisted Intervention--MICCAI 2016: 19th International Conference,
  Athens, Greece, October 17-21, 2016, Proceedings, Part II 19}},
  \bibinfo{pages}{424--432} (\bibinfo{organization}{Springer},
  \bibinfo{year}{2016}).

\bibitem{milletari2016v}
\bibinfo{author}{Milletari, F.}, \bibinfo{author}{Navab, N.} \&
  \bibinfo{author}{Ahmadi, S.-A.}
\newblock \bibinfo{title}{V-net: Fully convolutional neural networks for
  volumetric medical image segmentation}.
\newblock In \emph{\bibinfo{booktitle}{2016 fourth international conference on
  3D vision (3DV)}}, \bibinfo{pages}{565--571} (\bibinfo{organization}{Ieee},
  \bibinfo{year}{2016}).

\bibitem{simpson2019large}
\bibinfo{author}{Simpson, A.~L.} \emph{et~al.}
\newblock \bibinfo{journal}{\bibinfo{title}{A large annotated medical image
  dataset for the development and evaluation of segmentation algorithms}}.
\newblock {\emph{\JournalTitle{arXiv preprint arXiv:1902.09063}}}
  (\bibinfo{year}{2019}).

\bibitem{landman2015miccai}
\bibinfo{author}{Landman, B.} \emph{et~al.}
\newblock \bibinfo{title}{Miccai multi-atlas labeling beyond the cranial
  vault--workshop and challenge}.
\newblock In \emph{\bibinfo{booktitle}{Proc. MICCAI Multi-Atlas Labeling Beyond
  Cranial Vault—Workshop Challenge}}, vol.~\bibinfo{volume}{5},
  \bibinfo{pages}{12} (\bibinfo{year}{2015}).

\bibitem{antonelli2022medical}
\bibinfo{author}{Antonelli, M.} \emph{et~al.}
\newblock \bibinfo{journal}{\bibinfo{title}{The medical segmentation
  decathlon}}.
\newblock {\emph{\JournalTitle{Nature communications}}}
  \textbf{\bibinfo{volume}{13}}, \bibinfo{pages}{4128} (\bibinfo{year}{2022}).

\bibitem{zhang2023multi}
\bibinfo{author}{Zhang, M.} \emph{et~al.}
\newblock \bibinfo{journal}{\bibinfo{title}{Multi-site, multi-domain airway
  tree modeling (atm'22): A public benchmark for pulmonary airway
  segmentation}}.
\newblock {\emph{\JournalTitle{arXiv preprint arXiv:2303.05745}}}
  (\bibinfo{year}{2023}).

\bibitem{sundar2022fully}
\bibinfo{author}{Sundar, L. K.~S.} \emph{et~al.}
\newblock \bibinfo{journal}{\bibinfo{title}{Fully automated, semantic
  segmentation of whole-body 18f-fdg pet/ct images based on data-centric
  artificial intelligence}}.
\newblock {\emph{\JournalTitle{Journal of Nuclear Medicine}}}
  \textbf{\bibinfo{volume}{63}}, \bibinfo{pages}{1941--1948}
  (\bibinfo{year}{2022}).

\bibitem{hammers2003three}
\bibinfo{author}{Hammers, A.} \emph{et~al.}
\newblock \bibinfo{journal}{\bibinfo{title}{Three-dimensional maximum
  probability atlas of the human brain, with particular reference to the
  temporal lobe}}.
\newblock {\emph{\JournalTitle{Human brain mapping}}}
  \textbf{\bibinfo{volume}{19}}, \bibinfo{pages}{224--247}
  (\bibinfo{year}{2003}).

\bibitem{wasserthal2022totalsegmentator}
\bibinfo{author}{Wasserthal, J.} \emph{et~al.}
\newblock \bibinfo{journal}{\bibinfo{title}{Totalsegmentator: robust
  segmentation of 104 anatomical structures in ct images}}.
\newblock {\emph{\JournalTitle{arXiv preprint arXiv:2208.05868}}}
  (\bibinfo{year}{2022}).

\bibitem{xie2020self}
\bibinfo{author}{Xie, Q.}, \bibinfo{author}{Luong, M.-T.},
  \bibinfo{author}{Hovy, E.} \& \bibinfo{author}{Le, Q.~V.}
\newblock \bibinfo{title}{Self-training with noisy student improves imagenet
  classification}.
\newblock In \emph{\bibinfo{booktitle}{Proceedings of the IEEE/CVF conference
  on computer vision and pattern recognition}}, \bibinfo{pages}{10687--10698}
  (\bibinfo{year}{2020}).

\bibitem{zhai2022scaling}
\bibinfo{author}{Zhai, X.}, \bibinfo{author}{Kolesnikov, A.},
  \bibinfo{author}{Houlsby, N.} \& \bibinfo{author}{Beyer, L.}
\newblock \bibinfo{title}{Scaling vision transformers}.
\newblock In \emph{\bibinfo{booktitle}{Proceedings of the IEEE/CVF Conference
  on Computer Vision and Pattern Recognition}}, \bibinfo{pages}{12104--12113}
  (\bibinfo{year}{2022}).

\bibitem{kirillov2023segment}
\bibinfo{author}{Kirillov, A.} \emph{et~al.}
\newblock \bibinfo{journal}{\bibinfo{title}{Segment anything}}.
\newblock {\emph{\JournalTitle{arXiv preprint arXiv:2304.02643}}}
  (\bibinfo{year}{2023}).

\bibitem{reiss2023decoupled}
\bibinfo{author}{Rei{\ss}, S.}, \bibinfo{author}{Seibold, C.},
  \bibinfo{author}{Freytag, A.}, \bibinfo{author}{Rodner, E.} \&
  \bibinfo{author}{Stiefelhagen, R.}
\newblock \bibinfo{title}{Decoupled semantic prototypes enable learning from
  arbitrary annotation types for semi-weakly segmentation in expert-driven
  domains}.
\newblock In \emph{\bibinfo{booktitle}{Accepted to IEEE/CVF conference on
  computer vision and pattern recognition}} (\bibinfo{year}{2023}).

\bibitem{wang2022omni}
\bibinfo{author}{Wang, P.} \emph{et~al.}
\newblock \bibinfo{title}{Omni-detr: Omni-supervised object detection with
  transformers}.
\newblock In \emph{\bibinfo{booktitle}{Proceedings of the IEEE/CVF conference
  on computer vision and pattern recognition}}, \bibinfo{pages}{9367--9376}
  (\bibinfo{year}{2022}).

\bibitem{bae2022one}
\bibinfo{author}{Bae, W.}, \bibinfo{author}{Noh, J.},
  \bibinfo{author}{Asadabadi, M.~J.} \& \bibinfo{author}{Sutherland, D.~J.}
\newblock \bibinfo{journal}{\bibinfo{title}{One weird trick to improve your
  semi-weakly supervised semantic segmentation model}}.
\newblock {\emph{\JournalTitle{arXiv preprint arXiv:2205.01233}}}
  (\bibinfo{year}{2022}).

\bibitem{mlynarski2019deep}
\bibinfo{author}{Mlynarski, P.}, \bibinfo{author}{Delingette, H.},
  \bibinfo{author}{Criminisi, A.} \& \bibinfo{author}{Ayache, N.}
\newblock \bibinfo{journal}{\bibinfo{title}{Deep learning with mixed
  supervision for brain tumor segmentation}}.
\newblock {\emph{\JournalTitle{Journal of Medical Imaging}}}
  \textbf{\bibinfo{volume}{6}}, \bibinfo{pages}{034002--034002}
  (\bibinfo{year}{2019}).

\bibitem{qu2023annotating}
\bibinfo{author}{Qu, C.} \emph{et~al.}
\newblock \bibinfo{journal}{\bibinfo{title}{Annotating 8,000 abdominal ct
  volumes for multi-organ segmentation in three weeks}}.
\newblock {\emph{\JournalTitle{arXiv preprint arXiv:2305.09666}}}
  (\bibinfo{year}{2023}).

\bibitem{hu2023label}
\bibinfo{author}{Hu, Q.} \emph{et~al.}
\newblock \bibinfo{title}{Label-free liver tumor segmentation}.
\newblock In \emph{\bibinfo{booktitle}{Proceedings of the IEEE/CVF Conference
  on Computer Vision and Pattern Recognition}}, \bibinfo{pages}{7422--7432}
  (\bibinfo{year}{2023}).

\bibitem{slicer}
\bibinfo{author}{Kikinis, R.}, \bibinfo{author}{Pieper, S.~D.} \&
  \bibinfo{author}{Vosburgh, K.~G.}
\newblock \bibinfo{title}{3d slicer: a platform for subject-specific image
  analysis, visualization, and clinical support}.
\newblock In \emph{\bibinfo{booktitle}{Intraoperative imaging and image-guided
  therapy}}, \bibinfo{pages}{277--289} (\bibinfo{publisher}{Springer},
  \bibinfo{year}{2013}).

\bibitem{gatidis2022whole}
\bibinfo{author}{Gatidis, S.} \emph{et~al.}
\newblock \bibinfo{journal}{\bibinfo{title}{A whole-body fdg-pet/ct dataset
  with manually annotated tumor lesions}}.
\newblock {\emph{\JournalTitle{Scientific Data}}} \textbf{\bibinfo{volume}{9}},
  \bibinfo{pages}{601} (\bibinfo{year}{2022}).

\bibitem{xue2021multi}
\bibinfo{author}{Xue, Z.} \emph{et~al.}
\newblock \bibinfo{journal}{\bibinfo{title}{Multi-modal co-learning for liver
  lesion segmentation on pet-ct images}}.
\newblock {\emph{\JournalTitle{IEEE Transactions on Medical Imaging}}}
  \textbf{\bibinfo{volume}{40}}, \bibinfo{pages}{3531--3542}
  (\bibinfo{year}{2021}).

\bibitem{marinov2023mirror}
\bibinfo{author}{Marinov, Z.}, \bibinfo{author}{Rei{\ss}, S.},
  \bibinfo{author}{Kersting, D.}, \bibinfo{author}{Kleesiek, J.} \&
  \bibinfo{author}{Stiefelhagen, R.}
\newblock \bibinfo{journal}{\bibinfo{title}{Mirror u-net: Marrying multimodal
  fission with multi-task learning for semantic segmentation in medical
  imaging}}.
\newblock {\emph{\JournalTitle{arXiv preprint arXiv:2303.07126}}}
  (\bibinfo{year}{2023}).

\bibitem{isensee2021nnu}
\bibinfo{author}{Isensee, F.}, \bibinfo{author}{Jaeger, P.~F.},
  \bibinfo{author}{Kohl, S.~A.}, \bibinfo{author}{Petersen, J.} \&
  \bibinfo{author}{Maier-Hein, K.~H.}
\newblock \bibinfo{journal}{\bibinfo{title}{nnu-net: a self-configuring method
  for deep learning-based biomedical image segmentation}}.
\newblock {\emph{\JournalTitle{Nature methods}}} \textbf{\bibinfo{volume}{18}},
  \bibinfo{pages}{203--211} (\bibinfo{year}{2021}).

\bibitem{jordan2022pediatric}
\bibinfo{author}{Jordan, P.} \emph{et~al.}
\newblock \bibinfo{journal}{\bibinfo{title}{Pediatric chest-abdomen-pelvis and
  abdomen-pelvis ct images with expert organ contours}}.
\newblock {\emph{\JournalTitle{Medical Physics}}}
  \textbf{\bibinfo{volume}{49}}, \bibinfo{pages}{3523--3528}
  (\bibinfo{year}{2022}).

\bibitem{sekuboyina2021verse}
\bibinfo{author}{Sekuboyina, A.} \emph{et~al.}
\newblock \bibinfo{journal}{\bibinfo{title}{Verse: A vertebrae labelling and
  segmentation benchmark for multi-detector ct images}}.
\newblock {\emph{\JournalTitle{Medical image analysis}}}
  \textbf{\bibinfo{volume}{73}}, \bibinfo{pages}{102166}
  (\bibinfo{year}{2021}).

\bibitem{yang2021ribseg}
\bibinfo{author}{Yang, J.}, \bibinfo{author}{Gu, S.}, \bibinfo{author}{Wei,
  D.}, \bibinfo{author}{Pfister, H.} \& \bibinfo{author}{Ni, B.}
\newblock \bibinfo{title}{Ribseg dataset and strong point cloud baselines for
  rib segmentation from ct scans}.
\newblock In \emph{\bibinfo{booktitle}{Medical Image Computing and Computer
  Assisted Intervention--MICCAI 2021: 24th International Conference,
  Strasbourg, France, September 27--October 1, 2021, Proceedings, Part I 24}},
  \bibinfo{pages}{611--621} (\bibinfo{organization}{Springer},
  \bibinfo{year}{2021}).

\bibitem{liu2021deep}
\bibinfo{author}{Liu, P.} \emph{et~al.}
\newblock \bibinfo{journal}{\bibinfo{title}{Deep learning to segment pelvic
  bones: large-scale ct datasets and baseline models}}.
\newblock {\emph{\JournalTitle{International Journal of Computer Assisted
  Radiology and Surgery}}} \textbf{\bibinfo{volume}{16}},
  \bibinfo{pages}{749--756} (\bibinfo{year}{2021}).

\bibitem{ji2022amos}
\bibinfo{author}{Ji, Y.} \emph{et~al.}
\newblock \bibinfo{journal}{\bibinfo{title}{Amos: A large-scale abdominal
  multi-organ benchmark for versatile medical image segmentation}}.
\newblock {\emph{\JournalTitle{arXiv preprint arXiv:2206.08023}}}
  (\bibinfo{year}{2022}).

\bibitem{kuanquan_wang_2022_6361906}
\bibinfo{author}{Wang, K.} \emph{et~al.}
\newblock \bibinfo{title}{Pulmonary artery segmentation challenge 2022},
  \url{10.5281/zenodo.6361906} (\bibinfo{year}{2022}).

\bibitem{lambert2020segthor}
\bibinfo{author}{Lambert, Z.}, \bibinfo{author}{Petitjean, C.},
  \bibinfo{author}{Dubray, B.} \& \bibinfo{author}{Kuan, S.}
\newblock \bibinfo{title}{Segthor: Segmentation of thoracic organs at risk in
  ct images}.
\newblock In \emph{\bibinfo{booktitle}{2020 Tenth International Conference on
  Image Processing Theory, Tools and Applications (IPTA)}},
  \bibinfo{pages}{1--6} (\bibinfo{organization}{IEEE}, \bibinfo{year}{2020}).

\bibitem{ma2021abdomenct}
\bibinfo{author}{Ma, J.} \emph{et~al.}
\newblock \bibinfo{journal}{\bibinfo{title}{Abdomenct-1k: Is abdominal organ
  segmentation a solved problem?}}
\newblock {\emph{\JournalTitle{IEEE Transactions on Pattern Analysis and
  Machine Intelligence}}} \textbf{\bibinfo{volume}{44}},
  \bibinfo{pages}{6695--6714} (\bibinfo{year}{2021}).

\bibitem{koitka2021fully}
\bibinfo{author}{Koitka, S.}, \bibinfo{author}{Kroll, L.},
  \bibinfo{author}{Malamutmann, E.}, \bibinfo{author}{Oezcelik, A.} \&
  \bibinfo{author}{Nensa, F.}
\newblock \bibinfo{journal}{\bibinfo{title}{Fully automated body composition
  analysis in routine ct imaging using 3d semantic segmentation convolutional
  neural networks}}.
\newblock {\emph{\JournalTitle{European radiology}}}
  \textbf{\bibinfo{volume}{31}}, \bibinfo{pages}{1795--1804}
  (\bibinfo{year}{2021}).

\bibitem{jin2020deep}
\bibinfo{author}{Jin, L.} \emph{et~al.}
\newblock \bibinfo{journal}{\bibinfo{title}{Deep-learning-assisted detection
  and segmentation of rib fractures from ct scans: Development and validation
  of fracnet}}.
\newblock {\emph{\JournalTitle{EBioMedicine}}} \textbf{\bibinfo{volume}{62}},
  \bibinfo{pages}{103106} (\bibinfo{year}{2020}).

\bibitem{giske2011local}
\bibinfo{author}{Giske, K.} \emph{et~al.}
\newblock \bibinfo{journal}{\bibinfo{title}{Local setup errors in image-guided
  radiotherapy for head and neck cancer patients immobilized with a custom-made
  device}}.
\newblock {\emph{\JournalTitle{International Journal of Radiation Oncology*
  Biology* Physics}}} \textbf{\bibinfo{volume}{80}}, \bibinfo{pages}{582--589}
  (\bibinfo{year}{2011}).

\bibitem{stoiber2017analyzing}
\bibinfo{author}{Stoiber, E.~M.} \emph{et~al.}
\newblock \bibinfo{journal}{\bibinfo{title}{Analyzing human decisions in igrt
  of head-and-neck cancer patients to teach image registration algorithms what
  experts know}}.
\newblock {\emph{\JournalTitle{Radiation Oncology}}}
  \textbf{\bibinfo{volume}{12}}, \bibinfo{pages}{1--7} (\bibinfo{year}{2017}).

\bibitem{bejarano2019longitudinal}
\bibinfo{author}{Bejarano, T.}, \bibinfo{author}{De~Ornelas-Couto, M.} \&
  \bibinfo{author}{Mihaylov, I.~B.}
\newblock \bibinfo{journal}{\bibinfo{title}{Longitudinal fan-beam computed
  tomography dataset for head-and-neck squamous cell carcinoma patients}}.
\newblock {\emph{\JournalTitle{Medical physics}}}
  \textbf{\bibinfo{volume}{46}}, \bibinfo{pages}{2526--2537}
  (\bibinfo{year}{2019}).

\bibitem{bejarano2018head}
\bibinfo{author}{Bejarano, T.}, \bibinfo{author}{De~Ornelas~Couto, M.} \&
  \bibinfo{author}{Mihaylov, I.~B.}
\newblock \bibinfo{journal}{\bibinfo{title}{Head-and-neck squamous cell
  carcinoma patients with ct taken during pre-treatment, mid-treatment, and
  post-treatment dataset}}.
\newblock {\emph{\JournalTitle{The Cancer Imaging Archive}}}
  \textbf{\bibinfo{volume}{10}}, \bibinfo{pages}{K9} (\bibinfo{year}{2018}).

\bibitem{clark2013cancer}
\bibinfo{author}{Clark, K.} \emph{et~al.}
\newblock \bibinfo{journal}{\bibinfo{title}{The cancer imaging archive (tcia):
  maintaining and operating a public information repository}}.
\newblock {\emph{\JournalTitle{Journal of digital imaging}}}
  \textbf{\bibinfo{volume}{26}}, \bibinfo{pages}{1045--1057}
  (\bibinfo{year}{2013}).

\bibitem{liu2020early}
\bibinfo{author}{Liu, S.}, \bibinfo{author}{Niles-Weed, J.},
  \bibinfo{author}{Razavian, N.} \& \bibinfo{author}{Fernandez-Granda, C.}
\newblock \bibinfo{journal}{\bibinfo{title}{Early-learning regularization
  prevents memorization of noisy labels}}.
\newblock {\emph{\JournalTitle{Advances in neural information processing
  systems}}} \textbf{\bibinfo{volume}{33}}, \bibinfo{pages}{20331--20342}
  (\bibinfo{year}{2020}).

\bibitem{hatamizadeh2022unetr}
\bibinfo{author}{Hatamizadeh, A.} \emph{et~al.}
\newblock \bibinfo{title}{Unetr: Transformers for 3d medical image
  segmentation}.
\newblock In \emph{\bibinfo{booktitle}{Proceedings of the IEEE/CVF winter
  conference on applications of computer vision}}, \bibinfo{pages}{574--584}
  (\bibinfo{year}{2022}).

\bibitem{keller2021right}
\bibinfo{author}{Keller, K.} \emph{et~al.}
\newblock \bibinfo{journal}{\bibinfo{title}{Right atrium size in the general
  population}}.
\newblock {\emph{\JournalTitle{Scientific reports}}}
  \textbf{\bibinfo{volume}{11}}, \bibinfo{pages}{22523} (\bibinfo{year}{2021}).

\end{thebibliography}
